\documentclass[a4paper,fleqn,usenatbib]{mnras}
    
\usepackage{newtxtext,newtxmath}

\usepackage[T1]{fontenc}
\usepackage{ae,aecompl}

\usepackage{graphicx}   
\usepackage{amsmath}    
\usepackage{multirow}

\usepackage{epstopdf}
\usepackage{ifthen}
\usepackage{xcolor}
\usepackage[utf8]{inputenc}
\usepackage{import}
\usepackage{xspace}

\newcommand{\figsfolder}{figs/}
\graphicspath{{\figsfolder}}
\newcommand{\scfig}{.45}
\newcommand{\fsky}{f_{\rm sky}}
\newcommand{\Nf}{{\rm N_f}}
\newcommand{\Nc}{{\rm N_c}}
\newcommand{\Nb}{{\rm N_{\rm bin}}}

\newcommand{\srun}{s_{\rm run}}

\newcommand{\betad}{\beta_{\rm d}}
\newcommand{\betas}{\beta_{\rm s}}
\newcommand{\adust}{A_{\rm d}}
\newcommand{\asyn}{A_{\rm s}}
\newcommand{\alphaEE}{\alpha^{\rm EE}}
\newcommand{\alphaBB}{\alpha^{\rm BB}}

\newcommand{\wmap}{\textit{WMAP}\xspace}
\newcommand{\planck}{\textit{Planck}\xspace}

\definecolor{orange}{rgb}{1,0.5,0}
\definecolor{darkred}{rgb}{.7,0,0.2}

\usepackage{ulem} 

\title[QUIJOTE-MFI polarized synchrotron power spectra]{QUIJOTE scientific results -- XII. Characterization of the polarized synchrotron emission at the power spectrum level in the QUIJOTE-MFI wide survey}

\author[Vansyngel et al.]{
F. Vansyngel,$^{1,2}$
J.~A. Rubi\~{n}o-Mart\'{\i}n,$^{1,2}$\thanks{E-mail: jose.alberto.rubino@iac.es}
P. de Ferra,$^{2}$
R.~B. Barreiro,$^{3}$
R.~T. G\'enova-Santos,$^{1,2}$
\newauthor
B. Casaponsa,$^{3}$
M. Fern\'andez-Torreiro,$^{1,2}$
D. Adak,$^{1,2}$
K. Aryan,$^{1,2}$
M. Ashdown,$^{4,5}$
J.~M. Casas,$^{11,12}$
\newauthor
F.~J. Casas,$^{3}$
R. Cepeda-Arroita,$^{1,2}$ 
A. Fasano,$^{1,2}$
D. Herranz,$^{3}$
E. Mart\'{\i}nez-Gonz\'alez,$^{3}$
F. Poidevin,$^{1,2}$
\newauthor
R. Rebolo,$^{1,2,7}$
B. Ruiz-Granados,$^{8,1,2}$
D. Tramonte,$^{9,1,2}$
P. Vielva,$^{3}$
R.~A. Watson$^{10}$
\\
$^{1}$Instituto de Astrof\'{\i}sica de Canarias, E-38205 La Laguna, Tenerife, Spain\\
$^{2}$Departamento de Astrof\'{\i}sica, Universidad de La Laguna (ULL),
E-38206 La Laguna, Tenerife, Spain\\
$^{3}$Instituto de F\'{\i}sica de Cantabria (IFCA), CSIC-Univ. de Cantabria, Avda. los
Castros, s/n, E-39005 Santander, Spain\\
$^{4}$Astrophysics Group, Cavendish Laboratory, University of Cambridge,  J J Thomson Avenue, Cambridge CB3 0HE, UK\\
$^{5}$Kavli Institute for Cosmology, University of Cambridge, Madingley Road, Cambridge CB3 0HA, UK\\
$^{6}$Universidad de Oviedo, Departamento de F\'{\i}sica, C. Federico Garc\'{\i}a Lorca 18, 33007 Oviedo, Spain\\
$^{7}$Consejo Superior de Investigaciones Cient\'{\i}ficas, E-28006 Madrid, Spain\\
$^{8}$Departamento de F\'{\i}sica. Facultad de Ciencias. Universidad de C\'ordoba. Campus de Rabanales, Edif. C2. Planta Baja.  E-14071 C\'ordoba, Spain.\\
$^{9}$Department of Physics, Xi'an Jiaotong--Liverpool University, 111 Ren’ai Road, Suzhou Dushu Lake Science and Education Innovation District,\\ Suzhou Industrial Park, Suzhou 215123, China\\
$^{10}$Jodrell Bank Centre for Astrophysics, Alan Turing Building,  Department of Physics and Astronomy, School of Natural Sciences, The  University of Manchester,\\ 
Oxford Road, Manchester M13 9PL, Manchester, UK\\
$^{11}$Universidad de Oviedo, Departamento de Matem\'{a}ticas, C. Federico Garc\'{i}a Lorca 18, 33007 Oviedo, Spain\\
$^{12}$Instituto Universitario de Ciencias y Tecnolog\'{i}as Espaciales de Asturias (ICTEA), C. Independencia 13, 33004 Oviedo, Spain\\
}

\date{Accepted 2026 October 21. Received 2026 October 21; in original form 2026 July 22}

\pubyear{2026}

\begin{document}
\label{firstpage}
\pagerange{\pageref{firstpage}--\pageref{lastpage}}
\maketitle

\begin{abstract}
We present an analysis of the polarized synchrotron emission at the power-spectrum level in the multipole range $30 \leq \ell < 130$, using the recent QUIJOTE-MFI wide survey (10--20\,GHz) in combination with \planck and \wmap data. The $EE$ and $BB$ power spectra are well described by a simple power-law model, with no evidence for the need of additional parameters. The best-fit slopes are $\alphaEE = -3.24 \pm 0.12$ and $\alphaBB = -3.81 \pm 0.34$ for $|b|>20^\circ$, hinting toward a steepening of the $BB$ spectrum relative to $EE$ at the $\sim1.6\sigma$ level. The synchrotron frequency spectral indices are found to be $\betas^{\rm EE} = -3.26 \pm 0.05$ and $\betas^{\rm BB} = -3.25 \pm 0.10$ for the same cut, in agreement with earlier real-space analyses. We find no evidence for spatial variations of $\betas$ across different Galactic latitude cuts, within current uncertainties. The synchrotron $BB/EE$ power ratio is approximately 0.2 over the considered multipole range, lower than typical values for thermal dust, and consistent with previous measurements. We also measure a dust–synchrotron spatial correlation in the range 0.1--0.25 over $30 \leq \ell < 130$.
Using these results, we estimate the level of synchrotron contamination in future CMB $B$-mode measurements, finding an average power in the considered sky masks equivalent to $r \simeq 0.003$ at 90\,GHz and $r \simeq 0.0002$ at 150\,GHz for the relevant angular scales. 
\end{abstract}

\begin{keywords}
cosmology: observations -- cosmic microwave background -- radio continuum: ISM
\end{keywords}


\section{Introduction}

Cosmic microwave background (CMB) intensity anisotropies have been central to the development of modern cosmology, providing key measurements that have contributed to establishing the standard cosmological model with remarkable accuracy \citep{1992ApJ...396L...1S, 2013ApJS..208...20B, Planck2018-i}. As observational sensitivity has steadily improved, the emphasis has increasingly shifted towards the study of CMB polarization anisotropies, as they enable searches for primordial B modes, whose detection would provide direct evidence for a stochastic background of gravitational waves generated during cosmic inflation \citep{Kamion1997, PhysRevD.55.1830, 2016ARA&A..54..227K}.
Over the past decade, significant progress has been made in constraining the amplitude of such signals, with the most stringent upper limits on the tensor-to-scalar ratio currently provided by the BICEP/Keck experiment \citep{BK2021}, further strengthened by the inclusion of \planck data \citep{Tristram2022}, to the level of $r < 0.032$ at 95 per cent confidence level.

Looking ahead, a new generation of ground-based observatories, including the Simons Observatory \citep{so_19} and QUBIC \citep{QUBIC2024} in the southern hemisphere, and QUIJOTE \citep{RubinoSPIE12}, LSPE \citep{LSPE} and GroundBIRD \citep{Groundbird,2024SPIE13102E..05T} in the northern one, together with dedicated space missions such as LiteBIRD \citep{Litebird2023} and PICO \citep{PICO} aim to improve sensitivity to primordial B-modes by more than an order of magnitude, placing unprecedented demands on instrumental performance, systematic control and foreground subtraction. This paper focuses on this third aspect.

The main diffuse polarized contaminants are synchrotron emission (which dominates at frequencies below $\sim 70$\,GHz) and thermal dust emission (which is the dominant mechanism at high frequencies) \citep{Akrami:2018mcd}. In addition to synchrotron, two additional foregrounds, anomalous microwave emission (AME) and free-free, emit at low frequencies. AME has been  detected in intensity, but no polarization has been measured up to date \citep{AME2012, Draine_2016, AME2018,  vidal2026probinganomalousmicrowaveemission}, with current upper limits below the level of one per cent \citep{W44,perseus_w43_rhooph}. In addition, theoretical models predict a polarization fraction not larger than a few percent in local specific observation conditions, and lower in the general case \citep{Draine_2016, 2026arXiv260426001L}. Regarding free-free emission, polarization has not been detected \citep[see e.g.][]{2016A&A...594A..25P}.
According to the concordance model, the peak of the CMB $B$-mode power spectrum -- or CMB $BB$ power spectrum -- is expected to be at intermediate angular scale around multipole $\ell\simeq 80$ \citep[e.g.][]{Lewis:1999bs}. In this range of scales, there is no frequency domain or sky region where the sum of the synchrotron and thermal dust foregrounds is subdominant with respect to a CMB B-mode signal
if it has an amplitude of $r = 10^{-3}$ \citep{Adam:2014bub,Krachmalnicoff:2018imw}.

Any robust determination of the tensor-to-scalar ratio, $r$, will require an accurate separation of the cosmological signal from these Galactic foreground emissions \citep{Remazeilles:2015hpa, Errard:2018ctl}. Recent studies have revealed increasing complexity in both the spectral and spatial properties of Galactic dust and synchrotron emissions, posing significant challenges for component separation \citep{ChoiPage2015, PlanckInt2017-L,  Krachmalnicoff:2018imw, Fuskeland2021, Weiland2022, MFIcompsep_pol}. This complexity is particularly relevant in the case of polarized synchrotron emission. Owing to limited sensitivity in the low-frequency channels of \planck and \wmap, polarized synchrotron emission cannot be robustly characterized at intermediate and high Galactic latitudes. As a result, the constraints on the spectral behavior of polarized synchrotron emission in these regions remain limited when using satellite data. In this context, dedicated observations at lower frequencies are required to complement existing satellite data and to improve foreground modeling to the level necessary for future B-mode polarization studies.

Recently, three radio surveys are providing key and complementary information in polarization: the S-PASS experiment, which observed the sky from the South hemisphere at 2.3\,GHz \citep{Carretti2019}; the C-Band All-Sky Survey (C-BASS) experiment, which will cover the full sky at 5\,GHz \citep{Jones:2018lvt}, and the QUIJOTE (Q-U-I JOint TEnerife) experiment, a multi-frequency microwave survey carried out at the Teide Observatory, covering the 10--40\,GHz frequency range with angular resolutions between 18 arcmin and one degree \citep{RubinoSPIE12}. 

Previous studies based on \wmap and \planck data seem to indicate that the polarized synchrotron $\betas$ has a mean value close to $-3.15$  \citep{Watts_2024,Fuskeland:2014eoa,Akrami:2018wkt}. 
The inclusion of S-PASS data further constrains $\betas$, leading to a mean index value of $\betas = -3.22 \pm 0.08$, found between 2.3 and 33\,GHz \citep{Krachmalnicoff:2018imw}.
QUIJOTE results published so far on the polarized synchrotron properties are mostly based on real-space component separation methods. Using a parametric approach on a pixel-by-pixel basis, \citet{MFIcompsep_pol} found an average value of the synchrotron spectral index of $\betas = -3.08 \pm 0.13$ at two-degree angular scales. This result is consistent with the average synchrotron spectral index measured between the MFI 11\,GHz and \wmap 23\,GHz channels of  $\betas = -3.09 \pm 0.14$ \citep{mfiwidesurvey}. More recently, \citet{semiblind} applied a semi-blind component separation method and obtained an average value of $\betas = -3.11 \pm 0.21$.
All these results hint towards spatial variations of the polarized synchrotron spectral index  larger than the values reported in current models by a factor of 2--3 \citep[e.g.][]{Krachmalnicoff:2018imw, MFIcompsep_pol, Thorne_2017}. Quantifying the spatial variability of $\betas$ is crucial to estimate the proper amount of synchrotron power at the CMB frequencies observed by current and forthcoming experiments, as intrinsic variability in spectral index leads to higher synchrotron residuals at CMB frequencies.

This paper is part of a series associated with the QUIJOTE MFI wide survey \citep{mfiwidesurvey}, which covers approximately 29,000\,deg$^{2}$ including most of the northern sky, in four frequency bands centered at $11.1$, $12.9$, $16.8$, and $18.8$\,GHz. The survey achieves an angular resolution of approximately one degree and sensitivities in linear polarization in the range 35--40\,$\mu$K per degree beam.
This range of QUIJOTE-MFI frequencies is sufficiently close to the \planck and \wmap bands to mitigate depolarization effects due to Faraday rotation, while at the same time extending the frequency coverage and providing a longer lever arm to constrain the synchrotron spectral energy distribution (SED).

The analysis carried out in this paper focuses on the characterization of the spectral properties of polarized synchrotron emission using harmonic space tools, following a similar methodology to the one used in \cite{ChoiPage2015, PlanckInt2017-L,  Krachmalnicoff:2018imw, Martire2022}.
By combining QUIJOTE-MFI, \planck and \wmap data, we investigate if the spatial variation of $\betas$ is significant enough to induce a departure of the synchrotron power spectrum SED ($E$- and $B$-modes) from a pure power law in harmonic space. 
The dust is the dominant polarized foreground above 70\,GHz, and continues to contribute significantly at lower frequencies \citep{Adam:2015wua}.
We also include the frequencies of the polarized channels of the \planck high frequency instrument (HFI) in order to better remove the dust from the data during the analysis, and for a better characterization of the synchrotron-dust correlation. 
In our analysis, we assume that both free-free and AME have negligible polarization fractions.

Since polarized dust and synchrotron emissions are present at all observed frequencies, their relative contributions to the signal must be inferred in order to estimate their properties. The separation of the components can be obtained with blind methods such as SMICA \citep{2008arXiv0803.1814C} or ILC \citep{Bennett:1992,10.1111/j.1365-2966.2011.19497.x}, or with parametric methods such as Commander \citep{Eriksen:2007mx} or B-SeCRET \citep{Hoz_2020}. In analyses based on spherical harmonics or power spectra, the usual way to deal with spatially varying SEDs is to assign more than one component to each physical emission.
A multi-component version of the SED can be obtained by expanding the SED with respect to its parameters that vary across the sky. In the assumption that the variation of the SED parameters over the sky is small, a small number of components is needed for a good enough description, i.e. within the noise levels achieved by current experiments \citep{Chluba:2017rtj}. In this paper, we will study the number of components necessary to describe both the synchrotron and the dust contained in the data.
Finally, it is worth noting that characterizing foregrounds at the power-spectrum level is important for two main reasons. First, it provides direct measurements of the key observables relevant for CMB $B$-mode studies. Second, harmonic-space analyses are largely insensitive to some of the systematic effects that affect real-space analyses of ground-based data, such as uncertainties in the monopole determination and the impact of large-scale filtering.

The paper is organized as follows. In Sec.~\ref{sec:data} we give a description of the data used for the analysis, focusing on the power spectra. In Sec.~\ref{sec:method} we describe the methodology, including the foreground models and the fitting procedure. In Sec.~\ref{sec:modelsel} we present the results on the number of components needed to describe the foregrounds. In Sec.~\ref{sec:results} we present the results on the properties of the polarized synchrotron. We finally discuss and conclude in Sec.~\ref{sec:ccl}. Supplementary results on the dust and alternative models are presented in the appendices.

\section{Maps and power spectra}
\label{sec:data}
In our analysis, we consider polarization data from three experiments: QUIJOTE-MFI, \wmap, and \planck. We exploit the broad frequency coverage to provide a strong lever arm for constraining the synchrotron spectral energy distribution, while the higher-frequency channels help to disentangle synchrotron and dust emission and to characterize their cross-correlation.
All maps use a HEALPix\footnote{\url{http://healpix.sourceforge.net}}  \citep{2005ApJ...622..759G} pixelization. 
Details of each survey, together with the data combinations adopted in our analysis, and the codes used for power spectra computations are presented in the following subsections.

\subsection{Maps}
\subsubsection{\wmap and \planck data}
\label{sec:extdata}

For consistency with the analyses of \citet{Krachmalnicoff:2018imw} and \citet{Martire2022}, we use only the 30\,GHz channel of the Planck Low Frequency Instrument (LFI). 
However, unlike in those previous studies, we include also the four Planck High Frequency Instrument (HFI) channels at 100, 143, 217, and 353\,GHz to improve the separation between synchrotron and dust emission and to investigate their cross-correlation properties. Our results are based on the \planck PR4 data release \citep[NPIPE,][]{npipe}. All the maps were downloaded from the Planck legacy archive\footnote{\url{http://pla.esac.esa.int/pla/\#home}}.
Ninety-nine (numbers 201st to the 299th)  of the end-to-end (E2E) noise simulations described in \cite{npipe} are used to estimate noise power spectra of the maps. 
A subset of all available E2E simulations was chosen to balance computational efficiency with statistical precision. With the chosen binning size of 20 multipoles (see \ref{sec:powspec}), 99 simulations yield a sample power spectrum that approximates the true noise power spectrum within 2\,\%, a margin that is negligible relative to the \planck signal-to-noise ratio.

As in \citet{Krachmalnicoff:2018imw}, we use the \wmap nine-year polarization maps of the two lowest bands, K and Ka, with central frequencies 23 and 33\,GHz \citep{2013ApJS..208...20B}. 
We combine the differential assembly (D/A) maps of the first four years and and the last five years to construct a 1--4 year map and a 5--9 year map.
These are then used to produce a half-difference map, which is subsequently rescaled to match the noise level of the full nominal maps.
These half maps are used to perform null tests and estimate the noise power spectra. All maps were downloaded from the LAMBDA website\footnote{\url{https://lambda.gsfc.nasa.gov}}. 

All maps are converted to units of antenna temperature using the so-called unit conversion factors, and they are also color corrected using an average SED for the synchrotron and dust components and the \texttt{fastcc} code\footnote{\url{https://github.com/mpeel/fastcc}} \citep{fastcc}. The corresponding factors are presented in Table~\ref{tab:uccc}. 

\begin{table*}
  \centering
  \caption{Color correction (CC) and unit conversion (UC) factors adopted in this paper. CC factors are computed using \texttt{fastcc} for an average synchrotron spectrum with $\betas=-3.2$ and a modified black-body for the dust with $\betad=1.55$ and $T_d=19.6$\,K. UC factors convert ${\rm K}_{\rm CMB}$ to ${\rm K}_{\rm RJ}$, and are computed consistently with the definition of CC. For QUIJOTE-MFI and LFI, they correspond to the monochromatic conversion at the central frequency. For HFI, they follow the corresponding convention including the band-pass shift correction. }\label{tab:uccc}
  \begin{tabular}{l|ccccccccc}
  \hline
   & MFI & MFI & WMAP & WMAP & LFI & HFI & HFI & HFI & HFI\\ 
  Frequency (GHz) & 11.1 & 12.9 & 22.8 & 33 & 28.4 & 100   & 143   & 217    & 353 \\
  \hline
  CC  & 0.964 & 0.997 & 0.946 & 0.969 & 0.999 & 0.993 & 0.993 & 0.924 & 0.907 \\ 
  UC  & 0.997 & 0.996 & 0.987 & 0.972 & 0.979 & 0.788 & 0.590 & 0.333 & 0.075 \\
  \hline  
  \end{tabular}
\end{table*}

\subsubsection{QUIJOTE data}
\label{sec:qjdata}

We use the public\footnote{QUIJOTE-MFI DR1: \url{doi.org/10.26698/quijote-mfi-dr1}} QUIJOTE-MFI wide survey DR1 maps in polarization at 11.1 and 12.9\,GHz \citep{mfiwidesurvey}. We do not use the 17 and 19\,GHz maps since they have a low signal to noise ratio in the diffuse regions and do not reduce the uncertainty on the target parameters of the paper. 

The public half-maps are used to estimate the noise power spectra at 11 and 13\,GHz. The half-difference of the two half-maps can be regarded as a single realization of the instrumental noise. However, by computing its power spectrum and averaging over wide ranges of multipoles, the resulting estimate provides a realistic approximation to the true noise power spectrum, as shown in Sect.~7 of \citet{mfiwidesurvey}.
Both maps are converted in units of antenna temperature and color corrected using the factors of Table~\ref{tab:uccc}.

\citet{mfiwidesurvey} also describes a post-processing filter applied to the MFI DR1 maps to remove radio-frequency interference present at constant declination (FDEC). As shown in that work, this filtering does not modify the instrument transfer function at low multipoles. In intensity, the sky signal is recovered to better than 1\,\% down to $\ell \approx 10$, while in polarization the accuracy remains within 1\,\% down to $\ell \approx 30$. As a conservative choice, the power-spectrum analysis presented in this paper is therefore restricted to $\ell \geq 30$, ensuring that no transfer-function correction is required. 
It is therefore particularly interesting to compare the average synchrotron spectral index derived from the present harmonic-space analysis with that obtained from real-space analyses, since, as shown by \citet{MFIcompsep_pol}, the FDEC filtering can bias the recovery of $\beta_{\rm s}$ in the latter over specific sky regions if the map zero levels are not properly restored.

\subsubsection{Dataset combinations}
\label{sec:datasets}

We consider three different combinations of the maps described in Sects.~\ref{sec:extdata} and \ref{sec:qjdata}, defined as follows:
\begin{enumerate}
\item \textit{WM+Pl}: includes the \wmap nine-year K and Ka band maps, together with the \planck full-mission maps at 30, 100, 143, 217, and 353\,GHz, for a total of $\Nf=7$ frequency channels;
\item \textit{WM+Pl+QJ}: extends the \textit{WM+Pl} set by including the QUIJOTE-MFI maps at 11 and 13\,GHz, resulting in a total of $\Nf=9$ frequency channels;
\item \textit{NULL}: includes the same $\Nf=9$ frequency channels, but using half-difference maps for \wmap and QUIJOTE-MFI, and one noise realization for each \planck frequency channel.
\end{enumerate}
The \textit{WM+Pl} and \textit{WM+Pl+QJ} combinations are used to quantify the impact of including QUIJOTE-MFI data in the analysis. The \textit{NULL} set is employed to perform visual null tests, in particular to assess whether the adopted prior probabilities are overly informative.

\subsection{Angular power spectra}
\label{sec:powspec}

\begin{figure}
\centerline{\includegraphics[scale=\scfig]{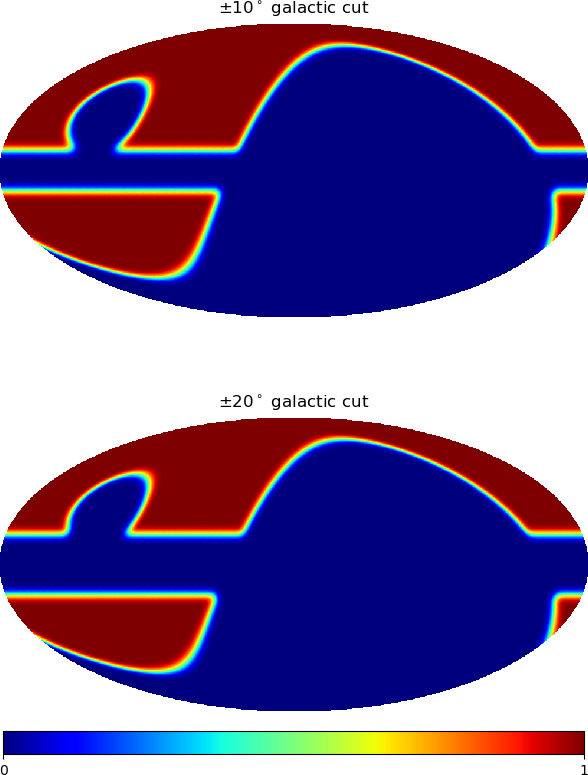}}
\caption{Example of two of the masks used in this analysis, corresponding to the combination of the QUIJOTE default analysis masks, and the galactic cuts $|b|>10^\circ$ (top) and $|b| >20^\circ$ (bottom). Here we consider three galactic cuts, namely 10, 15 and 20 degrees. The sky fractions after apodization are 34, 30, and 27~\%, respectively. See text for further details.}\label{fig:region}
\end{figure}

Our analysis focuses on high and intermediate Galactic latitudes, using three masks with different sky fractions. These masks are constructed by combining the default QUIJOTE analysis mask \citep[described in][and publicly available in the QUIJOTE-MFI DR1]{mfiwidesurvey} with Galactic latitude cuts that only include $|b| > b_{\rm cut}$, where $b_{\rm cut} = 10^\circ$, $15^\circ$, and $20^\circ$. In this way, we explicitly exclude a band around the Galactic plane, as this work is focused on diffuse emission regions.
For the power spectrum estimation, all masks are apodized using a Gaussian smoothing kernel with a full width at half maximum of $5^\circ$. After apodization, the resulting sky fractions are $\fsky = 34\%$, $30\%$, and $27\%$, respectively. The regions retained for the analysis are shown in red in Fig.~\ref{fig:region} for the two cases with the smallest and the largest sky coverage. 

We compute and use the auto-spectra of all frequencies and the cross-power spectra between all frequencies of the sets described in \ref{sec:datasets}. The spectra are computed using the PolSpice algorithm \citep{PolSpice}, which corrects the spectra for most of the mask effects. As an internal consistency check, we compared the spectra we obtained with PolSpice and those obtained with NaMaster \citep{NaMaster}, and we did not find any significant differences after binning the spectra. Loss of power due to the pixelization is compensated by dividing by the HEALPix pixel window function. The \wmap and \planck spectra are beam corrected using their respective instrument models. The QUIJOTE-MFI spectra are beam corrected using the beam transfer functions provided in their public data release.

We exclude from the analysis the larger scales of the maps that are dominated by highly non stationary and non Gaussian structures, which are complex to model in a power spectrum analysis. In addition, we limit the range of multipoles to keep the MFI data dominated regime. Thus, all spectra are binned in $\Nb=5$ groups of $\Delta\ell=20$ multipoles each, from $\ell=30$ to $\ell=129$, using the following weighted sum:
\begin{equation}\label{eq:bin}
C_b = \sum_\ell w_{b \ell} \, C_\ell \, ,
\end{equation}
where $b$ denotes the bin number, $C_\ell$ represents any auto- or cross-spectrum and $w_{b \ell} \propto (2\ell+1)$ for $\ell_{\rm min}(b) \leqslant \ell \leqslant \ell_{\rm max}(b)$ and $w_{b\ell} = 0$ elsewhere and, for each $b$, $\sum_\ell w_{b\ell} = 1$. This weighting scheme is motivated by the form of the binned likelihood used for the analysis (see Sec.~\ref{sec:methodfit} below). The central multipole of each bin is denoted as $\ell_b$. In the rest of the paper, a capital letter with $\ell$ as an index (as on the right-hand side of Eq.~\ref{eq:bin}) denotes an unbinned power spectrum and a capital letter with $b$ as an index (as on the left-hand side of Eq.~\ref{eq:bin}) denotes a binned power spectrum. Throughout the paper, angular power spectra in figures are  depicted with the usual band power $D_\ell = \ell (\ell +1) C_\ell / (2\pi)$. 

\begin{figure*}
\centerline{\includegraphics[scale=\scfig]{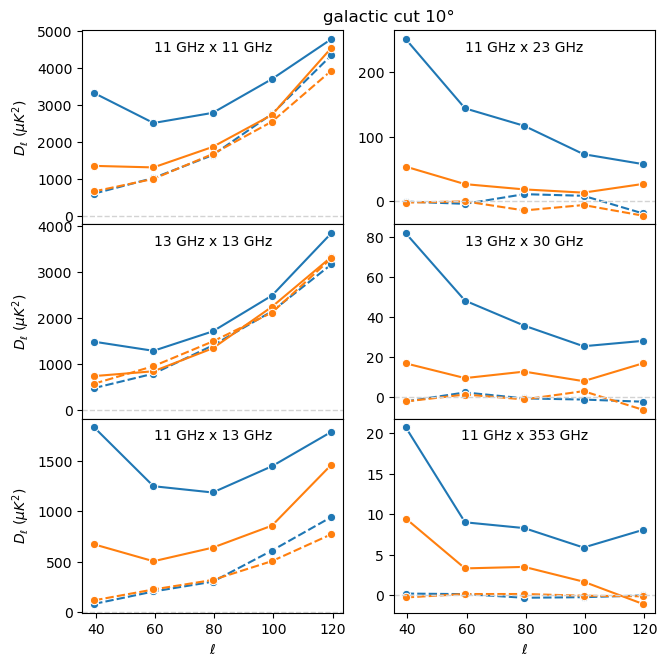}\hspace{.5cm}\includegraphics[scale=\scfig]{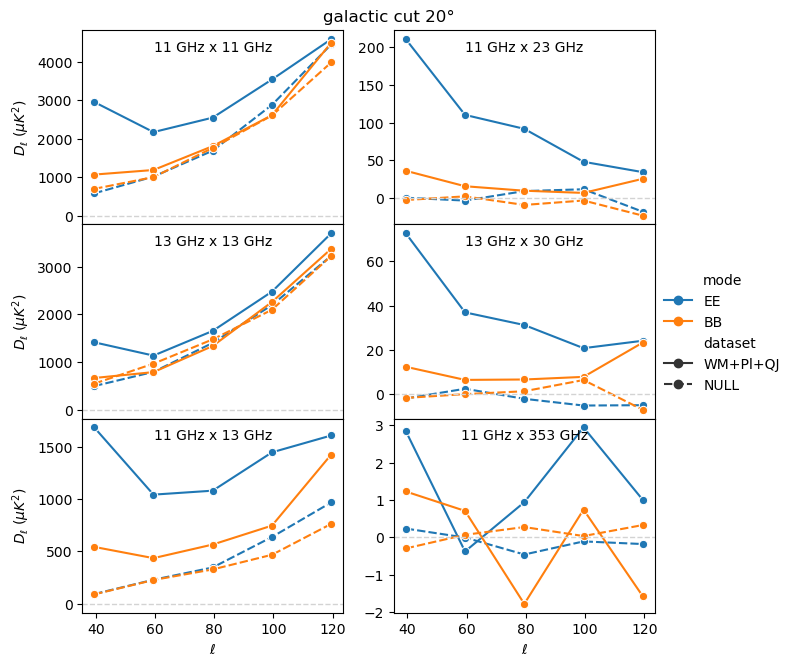}}
\caption{Selection of the auto- and cross-spectra used in the analysis. Left and right panels correspond to Galactic latitude cuts of 10$^\circ$ and 20$^\circ$, respectively. Within each panel, the left column shows the auto- and cross-spectra of the 11 and 13\,GHz channels of QUIJOTE-MFI Horn 3, while the right column shows the cross-spectra between QUIJOTE-MFI 11\,GHz and, from top to bottom, \wmap K-band (23\,GHz), \planck LFI (30\,GHz), and \planck HFI (353\,GHz). 
}\label{fig:spectra}
\end{figure*}

Fig.~\ref{fig:spectra} shows a selection of the auto- and cross-spectra ($D_\ell$) used for the analysis. Each panel displays the spectra for both the \textit{WM+Pl+QJ} and \textit{NULL} sets.
The \textit{NULL} set can be regarded as one realization of the instrumental noise, and the absolute values of the \textit{NULL} cross-spectra between QUĲOTE-MFI and \wmap or \planck provide an estimate of the typical amplitude of chance correlations between their instrumental noise realizations. For QUĲOTE-MFI data, the noise is correlated between 11 and 13\,GHz maps because both frequency channels of the same horn share a common amplifier. These correlations are accounted for in the noise power spectra and in the null test adopted in the analysis. The increase in power towards higher multipoles is driven by the $\ell(\ell+1)$ scaling factor in $D_\ell$  when applied to white-noise power spectra.
The cross-spectra between low-frequency maps show a clear synchrotron signal across all masks, particularly at low multipoles. 
Cross-correlating QUIJOTE-MFI with \planck and \wmap maps provides high signal-to-noise measurements of the polarized synchrotron emission. In particular, the 11\,GHz$\times$353\,GHz cross-spectra probe  the spatial correlation between synchrotron and thermal dust emission. 
We note that this correlation is substantially weaker for the 20$^\circ$ Galactic latitude cut than for the 10$^\circ$ cut, suggesting that, on the angular scales considered, it is largely driven by structures close to the Galactic plane. 

The $TT$, $TE$, $TB$, and $EB$ cross-spectra are not considered in the present paper, and $E$- and $B$-modes are treated separately.

\section{Method}
\label{sec:method}

This section describes the methodology adopted in this paper to estimate the foreground parameters. All analyses are performed in harmonic space, following an approach similar to that used in \citet{Akrami:2018mcd}, \citet{Krachmalnicoff:2018imw} and \citet{Martire2022}. The SEDs of the polarized foregrounds and the CMB are modeled using a parametric framework, and the E- and B-mode auto- and cross-power spectra are fitted to recover the key parameters describing the spectral behavior of synchrotron and dust emissions, their angular dependence, and their mutual cross-correlation.

\subsection{Foreground model}
\label{sec:modelfg}
We assume that the foregrounds consist of only polarized synchrotron and thermal dust emissions. As in previous works \citep[e.g.][]{Akrami:2018mcd, Krachmalnicoff:2018imw,Martire2022}, we assume that the level of linear polarization of free-free emission 
and AME is negligible \citep[e.g.][]{W44, AME2018,AMEHerman2023, perseus_w43_rhooph}. 

\subsubsection{Multi-component models}
\label{sec:sedexp}

For each of the two foreground emissions, namely the synchrotron or the dust, we consider two models. The first model is a single component model, i.e. the SED of the foreground does not vary spatially. This corresponds to a pure power law with spatially constant spectral index for the synchrotron, and a single modified blackbody with constant dust temperature and spectral index for the thermal dust emission. 
The second model is a two-component model where each component has a non spatially varying SED, but the sum of the two component results in one foreground that has a spatially varying SED. Taken separately, the two components of the same foreground emission do not necessarily have a physical meaning.

In the two-component model, the first component has the usual SED of the foreground, while for the second component we chose a first order correction term arising from possible small variations of the spectral indices of the foregrounds. The temperature of the dust is always supposed to be fixed to a value of $T_{\rm dust} = 19.6$\,K \citep{dustT}. 
These two components correspond to the first two terms in the SED moment expansion presented in \citet{Chluba:2017rtj}. This expansion allows to separate the spatial and spectral dependence of the foregrounds and treat one complex emission as several simpler components. Indeed, we can write the power-law frequency dependence $\nu^{\beta(\mathbf{n})}$, which is in principle different for each line of sight $\mathbf{n}$, as
\begin{equation}
\nu^{\beta(\mathbf{n})} = \nu^{\beta_0+\delta\beta(\mathbf{n})} \simeq \nu^{\beta_0} + \delta\beta(\mathbf{n})\nu^{\beta_0}\log\left(\nu\right) + \mathcal{O}\left(\delta\beta^2\right),
\end{equation}
where $\beta_0$ is the mean of the $\beta(\mathbf{n})$ map and $\delta\beta(\mathbf{n})$ is a zero-mean map. Thus, assuming $\delta\beta(\mathbf{n}) \ll 1$, we can approximate a line-of-sight dependent SED $a(\mathbf{n},\nu)$ of a map $s(\mathbf{n})$ as the sum of two independent maps $s_1$ and $s_2$ whose SED are fixed across the sky as follows
\begin{equation}\label{eq:decomp}
s(\mathbf{n})\, a(\mathbf{n},\nu) \simeq s_1(\mathbf{n})\, a_1(\nu) + s_2(\mathbf{n})\, a_2(\nu) \, .
\end{equation}
where $a_1(\nu)$ captures the pure power law dependence with constant spectral index, and $a_2(\nu)$ represents the spectral dependence of the first moment.
This decomposition can be applied to either total intensity maps, $E$- or $B$-mode maps.

\subsubsection{Sets of composite models}
\label{sec:modelsets}

As described in the previous subsection,
in this work we explore  
whether the data favors the single component description for synchrotron, dust or both, or the two-component model for any of the two emissions. 
The total number of synchrotron and dust components and the CMB (for $E$-modes) is denoted by $\Nc$. We can thus build four different models, depending on whether the foregrounds are each described by one or two components. A model selection will sort out the number of degrees of freedom needed to best describe the foregrounds. In what follows, the notation s$p$d$q$ refers to a model with a $p$-component synchrotron and a $q$-component dust (e.g. s2d1 stands for a 2 component synchrotron and 1 component dust and $\Nc=3$ without the CMB). As stated above, we only consider models with $p=1,2$ and $q=1,2$.

In the four possible s$p$d$q$ models, the values of the power spectra of the components are a priori independent from multipole bin to multipole bin, and thus are fitted separately in each bin. 
However, in addition to the four models, a model denoted by s$\alpha$d1 is used to fit a multipole spectral index for synchrotron, so we can compare with the literature values \citep{Krachmalnicoff:2018imw,Martire2022}. For this s$\alpha$d1 model, the dust is still a one-component dust with power spectra values independent from one bin to another, while the synchrotron is a one-component synchrotron but with a power spectrum parametrized as
\begin{equation}\label{eq:clindex}
S^{\rm synch.\times synch.}_\ell = \asyn \left(\frac{\ell}{80}\right)^{\alpha} ,
\end{equation}
using the same pivot scale of $\ell=80$ as in \citet{Krachmalnicoff:2018imw}. 
Then instead of fitting for the value of the power spectrum of each bin (like in the s1d1 model for example), only the overall amplitude $\asyn$ and the index $\alpha$ are fitted in this case.
As shown in \ref{sec:combined}, the amplitude parameter $\asyn$ is normalized at the pivot frequency of 23\,GHz. 

\subsection{CMB model}
\label{sec:modelcmb}

In this work, the CMB $E$-modes are fitted along with the foregrounds. Thus we have additional parameters corresponding to the amplitudes of the $EE$ power spectrum in each multipole bin. The SED of the CMB $E$-modes is kept fixed to the SED of the CMB anisotropies. We assume that the CMB does not contain any primordial $B$-modes, and that lensing-induced $B$-modes can be neglected as their amplitude is much lower than that of the total foreground emission in the multipole range studied in this work.

\subsection{Combined data model}
\label{sec:combined}

The data model used in this paper is a sum of two foregrounds (dust and synchrotron), CMB and noise. Thus for each multipole mode $(\ell,m)$ and for each polarization mode $E$ or $B$ (noted as $X$), the data vector can be decomposed as follows:
\begin{equation}\label{eq:datamod}
d_{\ell m}^X = A \, s_{\ell m}^X + n_{\ell m}^X \, .
\end{equation}
The piece of data $d_{\ell m}^X$ is a vector whose entries represent the $\Nf$ frequencies. The component vector $s_{\ell m}^X$ has size $\Nc$ and each entry represents either the CMB (for $EE$ only) or one component of the foregrounds, corresponding to the spherical harmonics transform of the $s_1$ and possibly $s_2$ maps of Eq.~\ref{eq:decomp}. The number of component $\Nc$ depends on the foreground model chosen among those presented in Sec.~\ref{sec:modelsets}. The $\Nf\times\Nc$ matrix $A$ is the mixing matrix and gathers the SEDs of the galactic emissions and of the CMB, in units of Rayleigh-Jeans temperature K$_{\rm RJ}$. Explicitly, the columns of the mixing matrix $A$ contain the following elements for the CMB:
\begin{equation}\label{eq:KCMB2KRJ}
a(\nu) = x_\nu^2 \exp(x_\nu) / (\exp(x_\nu)-1)^2,
\end{equation}
these for the synchrotron:
\begin{equation}
\begin{aligned}
a_1(\nu)&=(\nu/\nu_s)^{\betas}\\
a_2(\nu)&=(\nu/\nu_s)^{\betas}\log(\nu/\nu_s)
\end{aligned}
\end{equation}
and finally these for the thermal dust:
\begin{equation}\label{eq:SEDdustx2}
\begin{aligned}
a_1(\nu)&=(\nu/\nu_d)^{\betad+1}\,(\exp(y_{\nu_d})-1) / (\exp(y_\nu)-1)\\
  a_2(\nu)&=(\nu/\nu_d)^{\betad+1}\log(\nu/\nu_d)\,(\exp(y_{\nu_d})-1) / (\exp(y_\nu)-1), 
    \end{aligned}
\end{equation}
with $\nu_s=23\,{\rm GHz}$ and $\nu_d=353\,{\rm GHz}$ and where $x_\nu= h\,\nu/(k\,T_{\rm CMB})$, $y_\nu= h\,\nu/(k\,T_d)$, and $h$ and $k$ are the Planck and Boltzmann constants, respectively. 
The spectral dependence of the synchrotron is encoded in its spectral index $\betas$, while the spectral index for the thermal dust SED is given by $\betad$, with fixed dust temperature $T_d=19.6\,{\rm K}$.
For the synchrotron amplitude, we are using the pivot frequency of $\nu_s=23$\,GHz, while for the dust amplitude, we use $\nu_d=353$\,GHz. 

The vector $n_{\ell m}^X$ represents the instrumental noise in each channel and has same size as $d_{\ell m}^X$. The decomposition detailed in Sec.~\ref{sec:sedexp} allows to separate between frequency dependence (the mixing matrix) and the spatial dependence (the component vector). The power spectrum of the data model of Eq.~\ref{eq:datamod}, for either $E$- or $B$-modes, is then given by, after binning,
\begin{equation}\label{eq:covmod}
C_b(\theta) = A S_b A^T + N_b \, ,
\end{equation}
where superscript $T$ denotes the transposition, the covariance matrix $S_b$ gathers the binned power and cross spectra of the components and the matrix $N_b$ is the noise covariance. 

The correlation between any two components $i$ and $j$ among any of the synchrotron or dust components is $\rho_b^{i\times j}=S_b^{i\times j}/\sqrt{ S_b^{i\times i} S_b^{j\times j} }$. In the result section, only the correlation between the synchrotron and the dust of the s1d1 model is of interest and is simply noted $\rho_b$.

The diagonal elements of $N_b$ are computed using the maps described in Sects.~\ref{sec:extdata} and \ref{sec:qjdata}. 
All off-diagonal elements of $N_b$ are assumed to be zero except the covariance between the 11 and 13\,GHz channels of QUIJOTE-MFI, which is fixed to the cross power spectrum between the half-difference maps, as explained above.

Finally, the vector $\theta$ in Eq.~\ref{eq:covmod} is the set of the free parameters of the model, i.e. all entries of the binned signal covariance matrix $S_b$ (except the correlations between the CMB and the foregrounds that are kept fixed to zero), the mean power law indices $\betas$ and $\betad$ of the SED of the synchrotron and the dust emissions. The special case of the foreground model s$\alpha$d1 is detailed in the next subsection.

\subsection{Fitting procedure}
\label{sec:methodfit}

The method consists in building a posterior distribution over the set of model parameters $\theta$ and then sampling from the posterior possible values of $\theta$. The posterior is the product between a parametrized version of the SMICA likelihood \citep{2008arXiv0803.1814C} and some prior distribution over $\theta$. This method is a partial sky version of the blind Bayesian component separation presented in \citet{Vansyngel:2014dfa}. The inference is performed by sampling the posterior using the \texttt{stan} language \citep{stan} through the \texttt{rstan} interface \citep{rstan}.

The likelihood, for either $E$- or $B$-modes, is a Wishart distribution -- a generalization to matrices of the chi-squared and gamma distributions -- and can be written as follows
\begin{equation}\label{eq:likelihood}
\mathcal{L}(d\,\vert\, \theta) \propto \prod_{b} \left\vert C_b(\theta) \right\vert^{-\frac{k_b}{2}}\exp\left[-\frac{k_b}{2} {\rm Tr}\left\lbrace C_b(\theta)^{-1} \hat{C}_b(d) \right\rbrace\right]
\end{equation}
where the index $b$ runs over the number of multipole bins, ${\rm Tr}$ is the trace operation, $\hat{C}_b(d)$ and $C_b(\theta) = AS_bA^T+N_b$ are the binned power spectra of the data and of the data model respectively. The number $k_b$ is the number of degrees of freedom contained in bin $b$. We have $k_b = (2\ell_b+1)\Delta\ell\fsky$ where $\ell_b$ is the central multipole of bin $b$ \citep{Hivon_2002}.

The prior distributions on the dust and synchrotron emissivity indices are Gaussian distributions centered around the usual measured values: $\betas \thicksim \mathcal{N}(-3, 1)$ and $\betad \thicksim \mathcal{N}(1.6, 1)$.
The covariance of the signal $S_b$ can be separated into the CMB part, that is simply one entry for the CMB power spectrum, and the foreground part, $S_b^{{\rm fgd}}$. The covariance of the foregrounds $S_b^{{\rm fgd}}$ is reparametrized so that the sampled parameters are the square roots of the auto-spectra of the foregrounds $\sqrt{F_b}$ and the Cholesky factor of their Pearson correlation matrix $R_b = L_b L_b^T$, i.e. $S_b^{{\rm fgd}} =  \mathrm{diag}(\sqrt{F_b}) L_b L_b^T \mathrm{diag}(\sqrt{F_b})$. This reparametrization is meant to increase the efficiency of the sampling process. The prior distribution on $\sqrt{F_b}$ is a flat distribution (i.e. no prior distribution was used for $\sqrt{F_b}$) and the prior on $R_b$ is a uniform distribution, which in practice is implemented by using the Cholesky LKJ distribution of the \texttt{stan} language on $L_b$.
The prior on the CMB $E$-modes is very informative in order to mimic a removal of the \planck best fit \citep{Akrami:2018vks} from the data spectra and allowing some uncertainty in the process that will propagate to the uncertainties of all the other parameters. We choose an inverse-gamma distribution as the prior distribution for the CMB, since it is a natural choice to describe the prior of a variance. The parameters of the inverse-gamma prior distribution on the CMB $E$-modes are tuned so that the mode is the \planck 2018 best fit power spectrum $C_b^{{\rm Planck}}$ and that the variance is very close to the cosmic variance in a partial sky, such that $S_b^{\rm CMB} \thicksim \text{\Large $\gamma$}^{-1}\left(k_b,(k_b+1)C_b^{{\rm Planck}}\right)$ with $\text{\Large $\gamma$}^{-1}$ the probability distribution function of the inverse-gamma.

Regarding the s$\alpha$d1 model, the set of parameters $\theta$ of Eq.~\ref{eq:likelihood} does not contain the $N_b$ values of $\sqrt{F_b}$ for the synchrotron but instead contains the parameters $\asyn$ and $\alpha$ of Eq.~\ref{eq:clindex}. Both $\asyn$ and $\alpha$ have a flat prior distribution.

\section{Selection of the foreground model}
\label{sec:modelsel}

In this section we first select the model that best fits the data among those presented in Sec.~\ref{sec:datasets} and then we evaluate the goodness of fit of the best model.

\subsection{Model relative selection}
In order to select the most adapted foreground model among those presented in Sec.~\ref{sec:modelsets}, a model selection is performed using Bayes factors \citep{kass_bayes_1995}. The Bayes factor $B(M_i,M_j)$ provides a formal mechanism to compare two models, $M_i$ with set of parameters $\theta_i$ and $M_j$ with set of parameters $\theta_j$, by calculating the ratio of their marginal likelihoods (also known as evidence) as
\begin{equation}
B(M_i,M_j) = \frac{\int \mathcal{L}(d|\theta_i) P(\theta_i|M_i)\,\text{d}\theta_i}{\int \mathcal{L}(d|\theta_j) P(\theta_j|M_j)\,\text{d}\theta_j}
\end{equation}
with $P(\theta|M)$ the prior distribution over $\theta$ in model $M$. It is also important to note that penalty for more complex models is inherently implemented through integration over a larger number of parameters.

\begin{table}
\centering
\caption{Bayes factors in favor of the s$\alpha$d1 model against the other four models when using the dataset \textit{WM+Pl+QJ} on galactic cut $\pm10^\circ$, for $E$- and $B$-modes, expressed in base-10 logarithm, i.e.  $\log_{10}B({\rm s\alpha d1},M_j)$ with $M_j={\rm s1d1}, {\rm s2d1},\,{\rm s1d2},\,{\rm s2d2}$. The trend is very much similar on the two other galactic cuts and/or with the dataset \textit{WM+Pl}.}\label{tab:bf}
\begin{tabular}{ccccc}
  \hline
    & $M_j={\rm s1d1}$ &  $M_j={\rm s2d1}$  & $M_j={\rm s1d2}$ & $M_j={\rm s2d2}$\\
  \hline
  \hline
  EE & 5.13 & 11.9 & 16.5 & 23 \\ 
  BB & 4.58 & 10.5 & 11 & 15.9 \\ \hline
\end{tabular}
\end{table}

For each model, we perform an inference of the parameters using the data of the dataset \textit{WM+Pl+QJ}. Then we compute the evidence for each inference and the resulting Bayes factors between each pair of models with the R package \texttt{bridgesampling} \citep{bridgesampling}.
We consider the null hypothesis to be the simplest model we defined, namely the s$\alpha$d1 model. We use the Bayes factor to test whether the addition of further complexity presented in Sec.~\ref{sec:method} is supported by the data. Table~\ref{tab:bf} shows the ratios of the evidence of the s$\alpha$d1 model over the evidence of each of the other four models. 

Bayes factors can be interpreted as an odds ratio between the models. Following the guidelines suggested by \cite{kass_bayes_1995}, a Bayes factor $\log_{10}(B(M_i,M_j))>2$ is considered decisive evidence for model $M_i$, such that the Bayes factors of Table~\ref{tab:bf} indicate a significant preference for the s$\alpha$d1 model. Because Bayes factors are ratios of marginal likelihoods, and each entry in Table~\ref{tab:bf} uses the same reference model s$\alpha$d1, the values immediately determine the ranking of the four other models according to the evidence provided by the data. Two conclusions can be drawn from these values. First, the s1d1 model outperform the three other s$p$d$q$ models, meaning that only one component for the synchrotron and the dust is necessary to explain the data. In other words, the variability of the SED of the polarized synchrotron is not detected at the power spectrum level in the considered sky maps and for the data combination \textit{WM+Pl+QJ}. Second, the s$\alpha$d1 model is preferred over the s1d1 model, meaning that at the power spectrum level the data prefers the simplest considered model for the synchrotron, consisting in an angular power law spectrum and a frequency power law SED. The s$\alpha$d1 foreground model is therefore the model we will use in the remaining part of the paper to present the results of our analysis. For completeness, results for the second most favored model, s1d1, are also given in the Appendix~\ref{app:s1d1}.

\subsection{Goodness of fit}\label{sec:gof}

To quantitatively assess the goodness of fit, we first maximize the likelihood in Eq.~\ref{eq:likelihood} with respect to the parameters of the s$\alpha$d1 model and we note $\theta_m=\arg\underset{\theta}{\max}\mathcal{L}(d\,\vert\,\theta)$. The vector of parameters $\theta_m$ is then used to compute the log-likelihood ratio statistics
\begin{equation}\label{eq:llratio}
D = -2\log\left[\frac{\mathcal{L}(d\,\vert\,\theta_m)}{\mathcal{L}(d\,\vert\,\tau_m)}\right] = \sum_b D_b
\end{equation}
where $\tau_m$ is the set of parameters of the most general possible data model 
$\tau$ that maximizes the likelihood. In this model $\tau$, $C_b(\tau)$ is simply a $\Nf\times\Nf$ covariance without constraint and thus we have $C_b(\tau_m)=\hat{C}_b(d)$.
According to Wilks' theorem~\citep{wilks_large-sample_1938}, the quantities $D_b$ tend to be distributed as a chi-squared distribution with a number of degrees of freedom $n$, which is the number of degrees of freedom in a $\Nf$ covariance, i.e. $\Nf(\Nf+1)/2=45$, minus the number of parameters per multipole of the model. In the s$\alpha$d1 model there are
\begin{itemize}
    \item one dust power per multipole bin;
    \item one dust-synchrotron correlation per multipole bin;
    \item one synchrotron power spectrum amplitude scale for all multipole bins;
    \item one synchrotron power spectrum index scale for all multipole bins;
    \item one synchrotron SED index for all multipole bins;
    \item one dust SED index for all multipole bins;
\end{itemize}
i.e. there are $2+4/\Nb=2.8$ free parameters per multipole. 
The CMB is not counted as a free parameter, since it is very much constrained by the priors. Thus the number of degrees of freedom of the chi-squared distribution is $n=42.2$. We can then compute a goodness of fit metric from the log-likelihood ratios interpreted as chi-squared values. We choose the probability-to-exceed (PTE), which evaluates the probability of the occurrence of a larger deviation from the model of Eq.~\ref{eq:covmod} than what was observed in the data (e.g. \cite{pte}). In practice, PTE consists in evaluating the survival function of the chi-squared distribution of 42.2 degrees of freedom at the $D_b$ of Eq.~\ref{eq:llratio}.

Table~\ref{tab:pte} shows the PTE for the s$\alpha$d1 model in each multipole bin, separately for $EE$ and $BB$ modes. In the case of a perfect model and without correlation between the multipoles, the PTE values should appear as they were drawn from a uniform distribution. In general, the model provides a good fit to the $EE$ spectra. 
However, the first two bins of the $BB$ spectra appear to be poorly fitted. These low PTE values may be driven by the high signal-to-noise ratio at large angular scales, where the assumption of statistical isotropy breaks down and the dust and synchrotron emissions are non-stationary and non-Gaussian, in contrast with the assumptions underlying the SMICA likelihood. In addition, the associated uncertainty might be slightly underestimated in the likelihood for two reasons. First, the data spectra are computed on a small sky fraction, they are therefore different from the spectra computed on all possible patches with the same sky fraction. This variability is not taken into account in the likelihood. Second, the noise spectra of QUIJOTE-MFI and \wmap data are computed using half-difference maps, which would correspond to one particular realization of the noise and not the underlying true noise power spectrum.
Furthermore, the first bin contains less modes and is more prone to deviate from the chi-squared distribution compared to the other bins.
Finally, we note that the poor fit in the lowest multipole bins cannot be attributed to the assumed power-law dependence of synchrotron emission in the s$\alpha$d1 model. Indeed, we find that the PTE values obtained with the s1d1 model are similar to, or lower than, those of the s$\alpha$d1 model.

\begin{table}
\centering
\caption{PTE per mask, mode and multipole bin for the fit of the s$\alpha$d1 model to the \textit{WM+Pl+QJ} dataset.}\label{tab:pte}
\begin{tabular}{ccccccc}
\hline
Mask & Mode & 40 & 60 & 80 & 100 & 120 \\
\hline
\hline
 $\pm 10^\circ$ & $EE$ & 0.450 & 0.278 & 0.302 & 0.454 & 0.001 \\ 
 & $BB$ & $6\times 10^{-5}$ & 0.002 & 0.155 & 0.249 & 0.019 \\ 
\hline
$\pm 15^\circ$ & $EE$ & 0.114 & 0.192 & 0.648 & 0.335 & 0.006 \\ 
 & $BB$ & $4\times 10^{-5}$ & 0.024 & 0.538 & 0.148 & 0.079 \\
\hline 
$\pm 20^\circ$ & $EE$ & 0.145 & 0.137 & 0.737 & 0.186 & 0.024 \\ 
 & $BB$ & 0.002 & 0.024 & 0.752 & 0.172 & 0.071  \\
\hline
\end{tabular}
\end{table}

\section{Results}
\label{sec:results}

This section presents the results obtained when fitting the favored foreground model, namely the s$\alpha$d1 model. In the following plots, the points and error bars represent the median and the 16th to 84th percentile range of the marginal posterior distribution, corresponding to a 68\,\% confidence level, equivalent to $1\sigma$ for a normal distribution. Quantified results correspond to the median, with the $\pm$ values indicating the 16th to 84th interquantile difference, divided by 2. 
The spectral properties of polarized synchrotron emission are the primary focus of this paper. Accordingly, in this section we report only the synchrotron-related parameters, including the dust--synchrotron correlation. We note, however, that the methodology presented here also provides constraints on the thermal dust component. As a detailed discussion of the dust properties is beyond the scope of the present work, the corresponding results are presented in Appendix~\ref{app:dust_sfd1} for completeness.
In Appendix~\ref{app:s1d1} we present the parameters derived for the second-most favored model, s1d1, to illustrate that the key synchrotron parameters reported in this section (e.g. the spectral index) are robust with respect to the adopted model.
Finally, in Appendix~\ref{app:srun} we present the results of an alternative foreground model that includes the rolling of the synchrotron SED. 

In the following subsections, we first compare the data with the best-fitting model, and then present the inferred properties of the synchrotron emission, with particular emphasis on its role as a foreground for CMB studies.

\subsection{Data vs. model}

\newcommand{\scfigcen}{1.2}
\begin{figure*}
\centerline{\includegraphics[scale=\scfigcen]{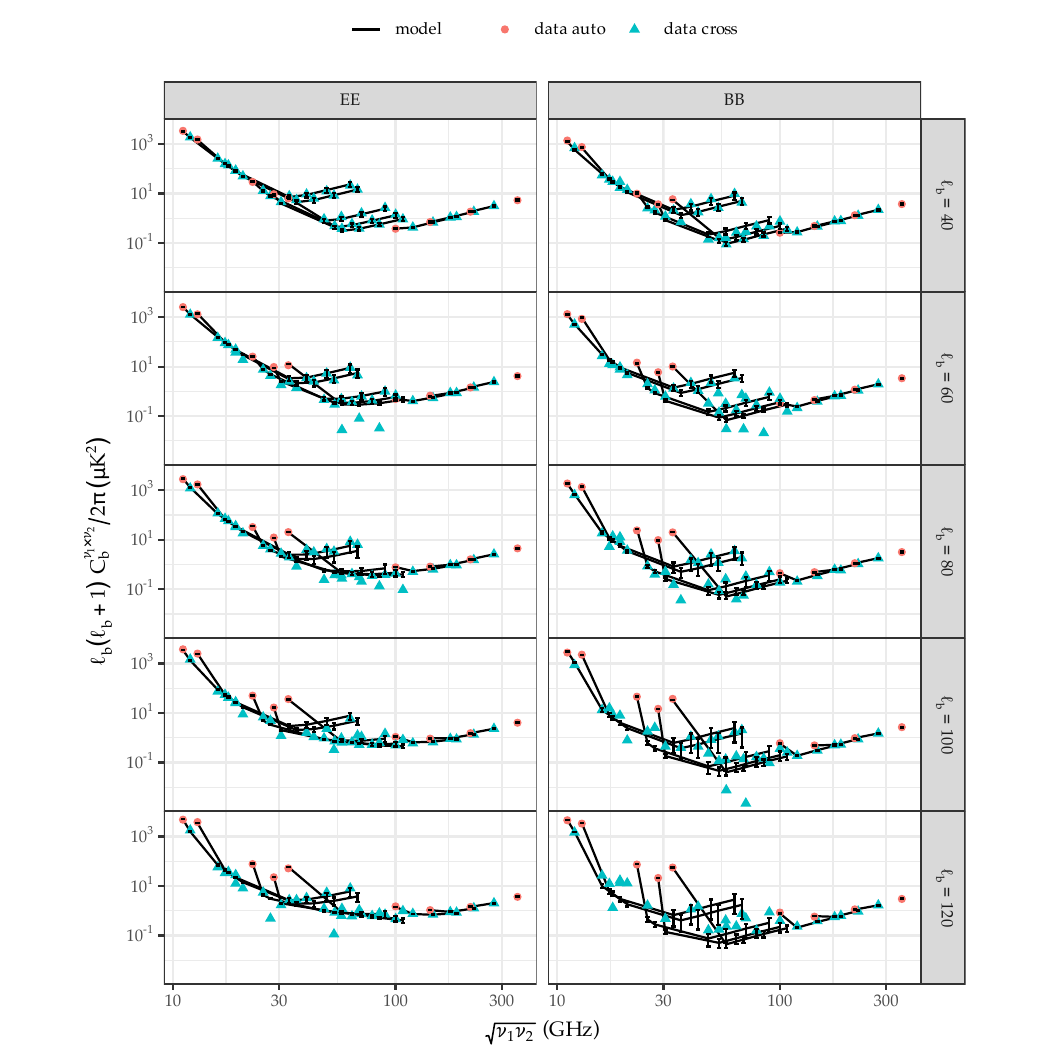}}
\caption{Power spectra from the \textit{WM+Pl+QJ} dataset using the $|b|>10^\circ$ Galactic mask. The auto- and cross-spectra are positioned along the x-axis according to their effective frequency, i.e. a spectrum between frequency $\nu_1$ and $\nu_2$ is positioned at the effective frequency $\sqrt{\nu_1\nu_2}$. The panels from top to bottom show the five multipole bins. The spectra are multiplied by a factor $\ell_b(\ell_b+1)/2\pi$ where $\ell_b$ is the central multipole of each bin. The best fit model is represented by the median values and the 15.9 to 84.1~\% quantiles. Each black line connects the set of best-fit spectra comprising an auto-spectrum at a given frequency and all its cross-spectra with higher frequencies. The red dots and the green triangles represent the values of the auto- and cross-spectra respectively. In most cases, the uncertainties lie within the line width. 
}\label{fig:datvsmod}
\end{figure*}

\begin{figure*}
\centerline{\includegraphics[scale=\scfigcen]{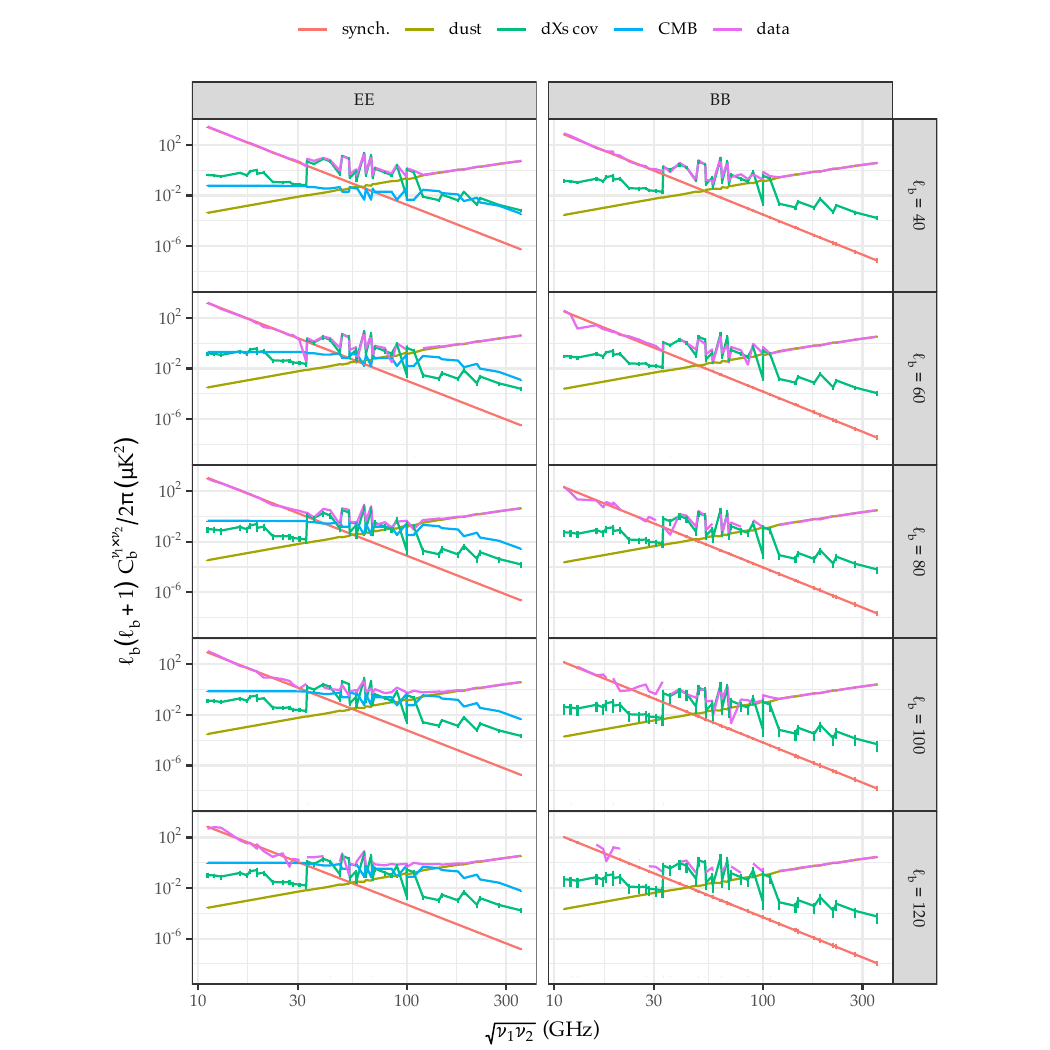}}
\caption{Best fit model. The layout of the auto- and cross-spectra and the panels is the same as in Fig.~\ref{fig:datvsmod}. The noise spectra have been subtracted from the data spectra. The figure shows the data spectra and the break down of the best-fit model into its individual components, namely synchrotron,  dust, the dust-synchrotron correlation (labeled as dXs) and the CMB. In most cases, the uncertainties lie within the line width.
}\label{fig:datvsmodcompo}
\end{figure*}

Fig.~\ref{fig:datvsmod} and~\ref{fig:datvsmodcompo} shows qualitatively the goodness of the fit by plotting the data cross- and auto-spectra for both the $EE$ and $BB$ cases, and the best fit data model for the $|b| < 10^\circ$ mask. Samples of the binned data model $C_\ell$ of Eq.~\ref{eq:covmod} are constructed from the samples of the parameters $\theta$. 
In both panels, the spectrum between two frequencies $\nu_1$ and $\nu_2$ is plotted against the effective frequency $\sqrt{\nu_1\nu_2}$. For a quantitative discussion on the goodness of fit, see Sec.~\ref{sec:gof}.
The measured power of the model can be decomposed into the synchrotron, the dust, the cross power between the synchrotron and the dust and the CMB contributions, as depicted in Fig.~\ref{fig:datvsmodcompo}. We note that if only the auto-spectra are taken into account, then the CMB is degenerate with the dust-synchrotron correlation at small scales. However, if the cross-spectra are added to the data, this degeneracy is broken and we do not observe any correlation between the CMB and the dust-synchrotron covariance. 

At the largest angular scales (i.e. the lowest multipole bins, which have the highest signal-to-noise ratio), the data and the model are in very good agreement, despite the low PTE values reported above. The covariance between synchrotron and dust emission, denoted as “dXs cov” in the plots, plays a major role in the fit, as nearly half of the measured cross-spectra are dominated by the dust–synchrotron correlation.

\subsection{Derived synchrotron properties}
\label{sec:synchprop}

\begin{table*}
\centering
\scriptsize
\caption{Values of the synchrotron related parameters when inferring using the s$\alpha$d1 model. The amplitude $A_s$ is computed at the pivot frequency of 23\,GHz. }
\begin{tabular}{cccccccccccc}
\hline
Mask & Dataset & Mode & $\asyn$ [$10^{-3}\mu K^2$] & $\alpha$ & corr($\asyn$,$\alpha$) & $\betas$ & $\rho_{40}$ & $\rho_{60}$ & $\rho_{80}$ & $\rho_{100}$ & $\rho_{120}$ \\
\hline
\hline
 $\pm 10^\circ$ & WM+Pl & EE & $9.26\pm0.59$ & $-3.10\pm0.12$ & 0.69 & $-3.00\pm0.09$ & $0.21\pm0.04$ & $0.11\pm0.04$ & $0.04\pm0.05$ & $0.18\pm0.05$ & $0.14\pm0.07$ \\ 
 &  & BB & $2.07\pm0.36$ & $-2.96\pm0.28$ & 0.86 & $-2.83\pm0.20$ & $0.15\pm0.05$ & $0.20\pm0.06$ & $0.08\pm0.08$ & $0.11\pm0.10$ & $0.27\pm0.13$ \\ 
 & WM+Pl+QJ & EE & $9.63\pm0.43$ & $-3.09\pm0.09$ & 0.54 & $-3.22\pm0.03$ & $0.20\pm0.04$ & $0.11\pm0.04$ & $0.08\pm0.04$ & $0.13\pm0.04$ & $0.12\pm0.05$ \\ 
 &  & BB & $1.52\pm0.22$ & $-3.54\pm0.22$ & 0.86 & $-3.33\pm0.07$ & $0.16\pm0.05$ & $0.17\pm0.05$ & $0.13\pm0.07$ & $0.13\pm0.09$ & $0.16\pm0.12$ \\ 
\hline
$\pm 15^\circ$ & WM+Pl & EE & $7.62\pm0.59$ & $-3.17\pm0.14$ & 0.74 & $-3.05\pm0.11$ & $0.09\pm0.05$ & $0.07\pm0.05$ & $0.05\pm0.05$ & $0.18\pm0.06$ & $0.09\pm0.08$ \\ 
 &  & BB & $1.47\pm0.36$ & $-3.29\pm0.37$ & 0.90 & $-3.01\pm0.25$ & $0.08\pm0.06$ & $0.19\pm0.07$ & $0.04\pm0.11$ & $0.15\pm0.13$ & $0.34\pm0.19$ \\ 
 & WM+Pl+QJ & EE & $8.09\pm0.42$ & $-3.13\pm0.10$ & 0.58 & $-3.24\pm0.03$ & $0.07\pm0.05$ & $0.07\pm0.04$ & $0.07\pm0.05$ & $0.11\pm0.05$ & $0.07\pm0.06$ \\ 
 &  & BB & $1.06\pm0.20$ & $-3.84\pm0.28$ & 0.89 & $-3.34\pm0.08$ & $0.09\pm0.05$ & $0.13\pm0.06$ & $0.03\pm0.09$ & $0.14\pm0.12$ & $0.17\pm0.17$ \\ 
\hline
$\pm 20^\circ$ & WM+Pl & EE & $6.81\pm0.61$ & $-3.26\pm0.16$ & 0.76 & $-3.12\pm0.12$ & $0.09\pm0.05$ & $-0.03\pm0.05$ & $0.02\pm0.06$ & $0.17\pm0.07$ & $0.12\pm0.09$ \\ 
 &  & BB & $1.39\pm0.39$ & $-3.18\pm0.43$ & 0.89 & $-3.01\pm0.30$ & $0.11\pm0.07$ & $0.19\pm0.08$ & $0.03\pm0.12$ & $0.10\pm0.15$ & $0.26\pm0.21$ \\ 
 & WM+Pl+QJ & EE & $7.09\pm0.43$ & $-3.24\pm0.12$ & 0.62 & $-3.26\pm0.03$ & $0.07\pm0.05$ & $-0.01\pm0.05$ & $0.04\pm0.05$ & $0.11\pm0.05$ & $0.04\pm0.07$ \\ 
 &  & BB & $0.93\pm0.22$ & $-3.81\pm0.34$ & 0.90 & $-3.25\pm0.10$ & $0.09\pm0.06$ & $0.16\pm0.07$ & $-0.02\pm0.10$ & $0.13\pm0.14$ & $0.12\pm0.20$  \\
\hline
\end{tabular}
\label{tab:sfd1}
\end{table*}

Table~\ref{tab:sfd1} shows the best-fit values of the synchrotron spectral parameters for the s$\alpha$d1 model, namely the amplitude at 23\,GHz ($\asyn$) and angular dependence ($\alpha$) of the synchrotron power spectrum as defined in equation~\ref{eq:clindex}; the synchrotron spectral index ($\betas$), and the synchrotron-dust correlation ($\rho_{b}$) for each one of the different multipole bins $b$. 
The s$\alpha$d1 model is preferred over the s1d1 model, as indicated by the Bayes factor (see Sec.~\ref{sec:modelsel}), but also because the distribution over all the other parameters of the model remains unchanged. We now discuss in detail these results.

\subsubsection{Amplitude and angular dependence ($\asyn, \alpha$)}\label{sec:alpha}

\begin{figure}
\centerline{\includegraphics[scale=\scfig]{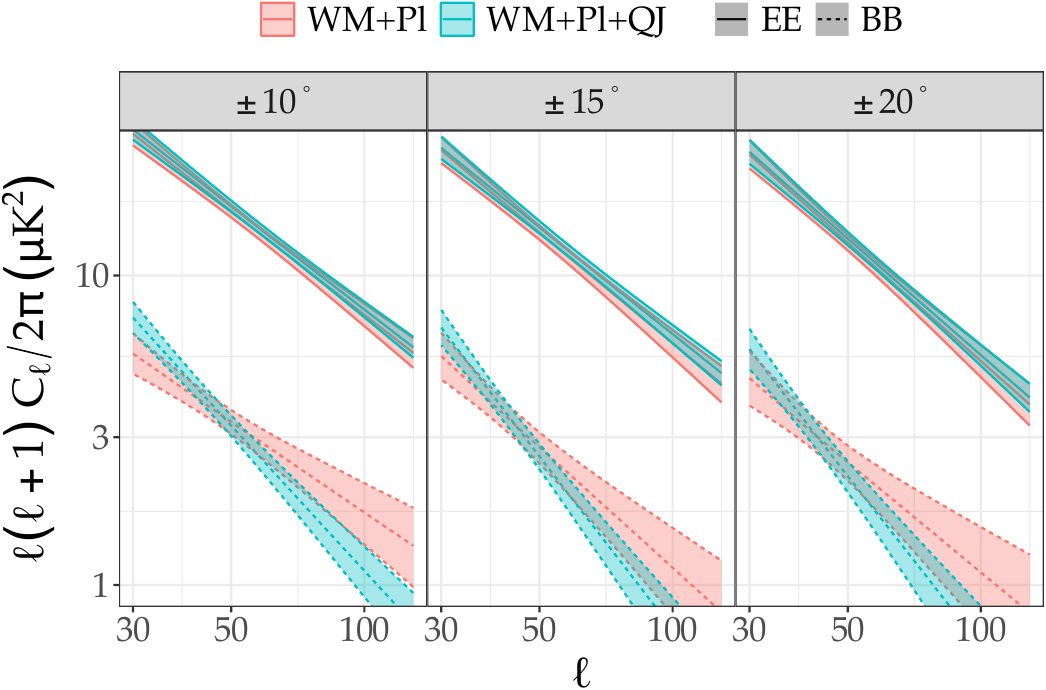}}
\caption{Best-fit synchrotron power spectra for the $E$- and $B$-modes at 23\,GHz, and for the two datasets \textit{WM+Pl} (red), \textit{WM+Pl+QJ} (blue). The line and shadow represent the median and the 15.9 to 84.1 \% quantiles. The three panels show, from left to right, the three analysis masks corresponding to $|b| < 10^\circ$, $15^\circ$ and $20^\circ$, respectively. 
}\label{fig:synchspec}
\end{figure}

\begin{figure}
\centerline{\includegraphics[scale=\scfig]{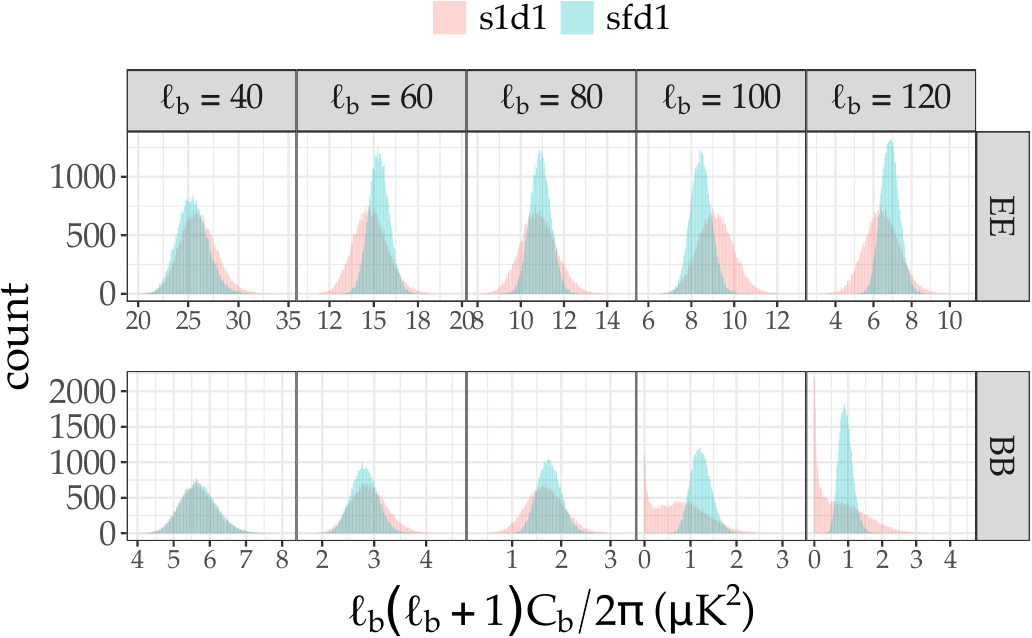}}
\caption{Marginal distributions of the binned power spectrum of the synchrotron. The green histograms are the distribution of the binned power spectra of the s$\alpha$d1 model, i.e. with parametrization by a power law, from the set \textit{WM+Pl+QJ} of Fig.~\ref{fig:synchspec}. The red histograms are the distributions of the binned power spectra in the s1d1 model, i.e. without parametrization by a power law. Without a parametric model for the synchrotron spectrum, the amplitudes at each multipole bins are almost independent and no synchrotron $BB$ signal is detected in the last two bins above $\ell = 80$.}\label{fig:Asdist}
\end{figure}

Fig.~\ref{fig:synchspec} shows the $EE$ and $BB$ synchrotron fitted spectra at 23\,GHz as inferred from the two datasets \textit{WM+Pl} and \textit{WM+Pl+QJ}.
The values of the amplitude $\asyn$ and index $\alpha$ are presented in Table~\ref{tab:sfd1}, for the three galactic masks, the $EE$ and $BB$ spectra separately and the two datasets \textit{WM+Pl+QJ} and \textit{WM+Pl}. The values in the table shows that QUIJOTE-MFI brings a significant amount of information about the synchrotron emission, reducing the error bars on $A_s$ by 27 to 44~\% and on $\alpha$ by 21 to 28~\%. The $EE$ spectrum is well constrained over the entire multipole range $30 \le \ell < 130$. The $BB$ spectrum is also well determined, but most of the information of the small scales is extrapolated from the large scales. 
This point is also illustrated in Fig.~\ref{fig:Asdist}, where we show that, without the parametrization of the synchrotron spectrum of Eq.~\ref{eq:clindex} (i.e. using the s1d1 model), the $BB$ spectrum of the synchrotron above multipole $\ell=80$ is poorly estimated. In the smallest scales of the s1d1 model, corresponding to the two highest multipole bins, the distribution of the spectrum in the \textit{NULL} set gives amplitudes similar to those in the other two sets.

The derived slopes ($\alpha$) of the $EE$ synchrotron power spectra in the range $30\le\ell<130$ and listed in Table~\ref{tab:sfd1} are in good agreement with those reported by \citet{Krachmalnicoff:2018imw}, based on S-PASS data over the multipole range $20<\ell<1000$, and with those of \citet{casas2026spatialvariationspolarizedsynchrotron}, based on all MFI channels over the multipole range $50<\ell<200$, although they are slightly steeper than those reported by \citet{Martire2022} for \planck and \wmap data over the range $30<\ell<300$. 
This discrepancy may be attributed to the different sky masks and multipole ranges considered, as well as to differences in the assumptions underlying the likelihood function.
Finally, when including QUIJOTE-MFI data in the analysis, we find marginal evidence for a steepening of the $\alphaBB$ values with respect to the $\alphaEE$ ones. The differences arises because the error bars are shrinking, but also because the value of $\alphaBB$ is different by 1.2-1.6$\sigma$ in the two datasets \textit{WM+Pl} and \textit{WM+Pl+QJ} whereas $\alphaEE$ has the same value in the two datasets. We find a $1.9$, $2.4$ and $1.6$ $\sigma$ difference between $\alphaEE$ and $\alphaBB$ on the $\pm10^\circ$, $\pm15^\circ$ and $\pm20^\circ$ masks, respectively. The first two multipole bins provide the majority of the information that constrains $\alpha$ (particularly for $BB$, as illustrated in Fig.~\ref{fig:Asdist}). However, these bins also exhibit the lowest PTE values. Consequently, the significance of the differences between the slopes of the $EE$ and $BB$ spectra must be put in perspective to the low statistical agreement between the model and the data at low multipoles.

\subsubsection{Synchrotron $B$-to-$E$ ratio}
\begin{figure}
\centerline{\includegraphics[scale=\scfig]{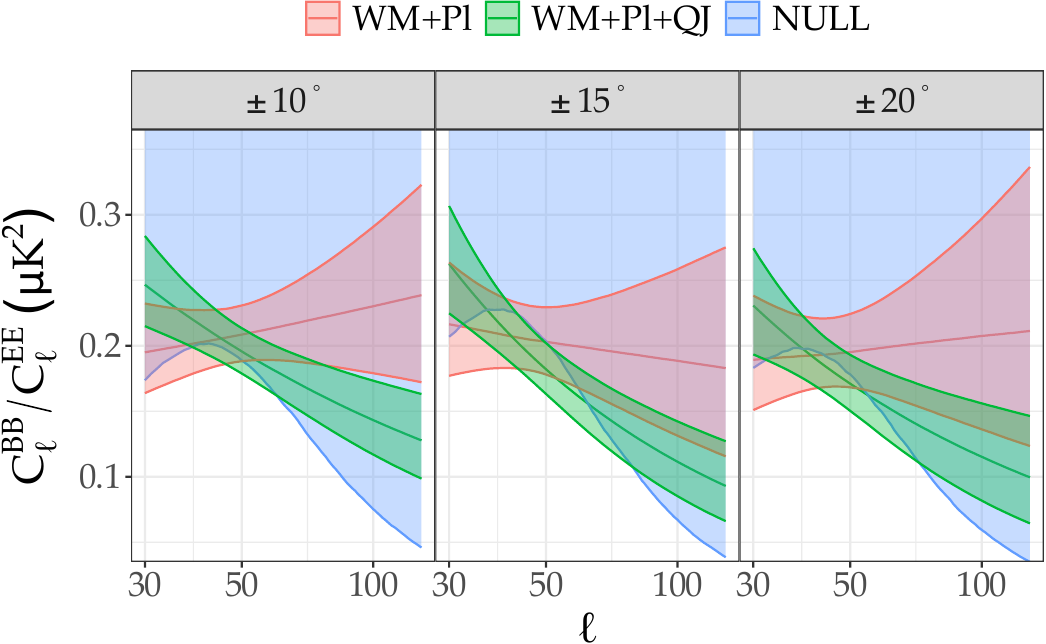}}
\caption{$BB$/$EE$ ratio of the synchrotron spectra at 23 GHz fitted on the three different datasets, and on the three galactic cuts.}\label{fig:synchratio}
\end{figure}

Fig.~\ref{fig:synchratio} shows the ratio between the $BB$ and $EE$ synchrotron spectra, as function of multipoles. Since the slope of the ratio is the difference between $\alphaBB$ and $\alphaEE$, the analysis of the results is similar to that of section~\ref{sec:alpha}. We find ratio values between $0.1$ and $0.25$, with a moderate dependence on the multipole when the combination \textit{WM+Pl+QJ} is used, thanks to the reduction of the error bars on the  power spectra. In the case of \textit{WM+Pl}, we do not find a clear dependence on scale, and the average value of the $B/E$ ratio is approximately $0.2$ for the three sky masks.
These numbers are consistent with the values reported by \citet{Martire2022} using \wmap K-band and \planck LFI30 data only, which range from $0.19\pm 0.01$ to $0.25\pm 0.02$ for their two extreme sky masks with sky fractions 94\,\% and 30\,\%, respectively; and also with the value reported in \citet{mfiwidesurvey} of $0.26\pm 0.07$ 
for $|b|>10^\circ$ using the QUIJOTE-MFI 11\,GHz map in the range $30 < \ell < 200$.
Therefore, we confirm that for synchrotron emission the B-to-E power spectrum ratio is of the order of $0.2$--$0.25$ when using QUIJOTE-MFI, \wmap and \planck data, in contrast to the values reported for thermal dust emission, which are typically of the order of $0.5$ \citep{Adam:2014bub, Planck2018-xi}. 

However, these $B$-to-$E$ ratios are significantly 
lower than what has been reported by  \citet{Krachmalnicoff:2018imw} using S-PASS data at 2.3\,GHz. 
Although the S-PASS and QUIJOTE-MFI sky footprints have a minimal overlap which might explain part of the difference \citep{casas2026spatialvariationspolarizedsynchrotron}, it is clear that the S-PASS results show a much stronger dependence with the Galactic latitude, with values ranging from $0.87\pm0.02$ for the sky mask including exclusively $|b|>20^\circ$, to $0.41\pm 0.08$ for $|b|>50^\circ$. As suggested in their paper, this is probably ascribed to some degree of Faraday rotation of the polarization angle of the signal at mid Galactic latitudes.

\subsubsection{Synchrotron spectral index $(\betas)$}

\begin{figure}
\centerline{\includegraphics[scale=\scfig]{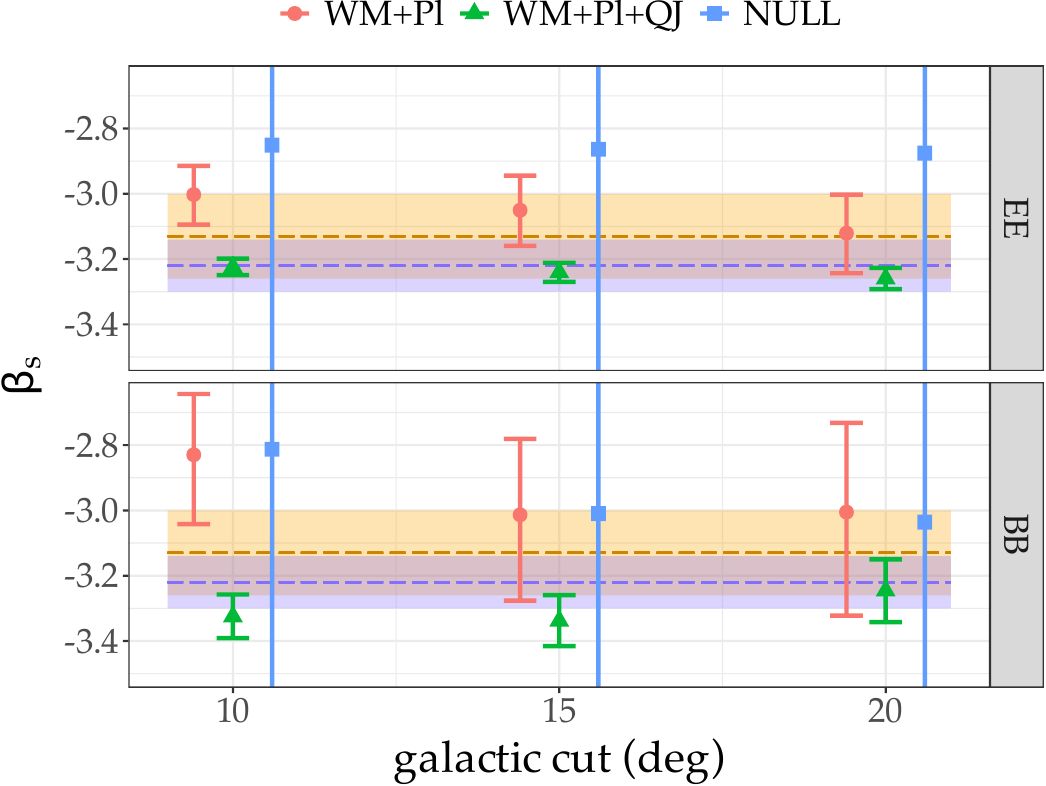}}
\caption{Synchrotron SED spectral index fitted on the three datasets, and on the three Galactic cuts. The shaded areas represent the \planck (\citet{Akrami:2018mcd} orange shade) and S-PASS values (\citet{Krachmalnicoff:2018imw} purple shade). Numerical values of $\betas$ can be found in Table~\ref{tab:sfd1}.}\label{fig:betas}
\end{figure}

Fig.~\ref{fig:betas} shows the synchrotron spectral indices $\betas$ obtained from the fits on the three datasets with the s$\alpha$d1 model. The numerical values are listed in Table~\ref{tab:sfd1} for the \textit{WM+Pl} and \textit{WM+Pl+QJ} datasets. The spectral index using the \textit{NULL} dataset is centered close to $-3$ because of the prior distribution adopted of $\betas \thicksim \mathcal{N}(-3, 1)$. The error bars in this \textit{NULL} case are much larger than those in the two other datasets, indicating that the data is the main source of information in the estimation of $\betas$ with both the \textit{WM+Pl} and \textit{WM+Pl+QJ} datasets. The addition of the QUIJOTE-MFI frequencies leads to the reduction of the error bars on $\betas$ by a factor 2 to 3. We do not find significant differences between the $EE$ and $BB$ spectral index. 

Our values are consistent with those works based on harmonic space analyses. \citet{Martire2022}  found $\betas^{\rm EE} = -2.95 \pm 0.10$ and $\betas^{\rm BB} = -3.02 \pm 0.35$ for the mask with $f_{\rm sky}=50$\,\%. 
Similarly, we also find consistency with the values reported by \citet{Krachmalnicoff:2018imw}, 
of $\betas=-3.22\pm0.08$ for the southern sky.  

The derived values are also consistent with studies based on real-space analyses of polarized synchrotron emission using QUIJOTE, \wmap, and \planck data. For example, \citet{MFIcompsep_pol} report a mean spectral index of $-3.08$ with a dispersion of $0.13$ using a parametric component separation method, while a semi-blind approach yields an inverse-variance-weighted mean synchrotron index of $-3.11$ with a standard deviation of $0.21$ \citep{semiblind}.

Finally, we note that the spectral indices recovered from the \textit{WM+Pl+QJ} and \textit{WM+Pl} datasets show a systematic difference, with the tension increasing with $\fsky$ and reaching $2.3\sigma$ for the 10-degree galactic cut where the signal-to-noise is larger. Although this is not yet statistically significant, it could hint toward a flattening of the synchrotron SED at higher frequencies, a potentially interesting result that warrants further investigation with future QUIJOTE-MFI2 data \citep[see e.g.][]{mfi2_synch_forecast}. 
To explore whether this apparent shift in the spectral index upon including the lower-frequency QUIJOTE-MFI data could be indicative of spectral running, in Appendix~\ref{app:srun} we repeat the analysis by introducing a running parameter into the s$\alpha$d1 model. Although the inferred running is consistent with zero, we find that the shift in $\beta_s$ is correlated with the statistical significance of the inferred running.

\subsubsection{Correlation matrix of synchrotron parameters}

\begin{figure}
\centerline{\includegraphics[scale=.45]{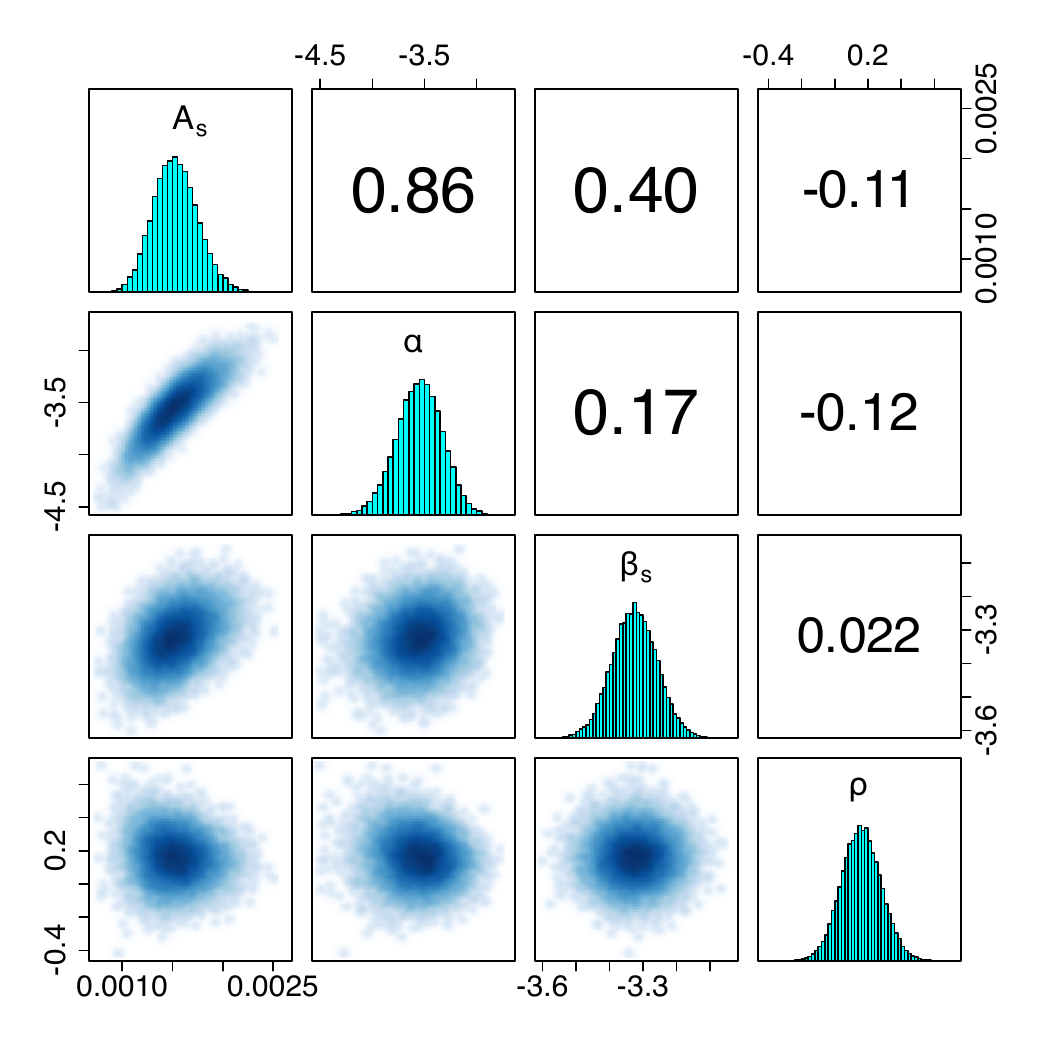}}
\caption{Example of the correlation structure of the four parameters describing the synchrotron component of the s$\alpha$d1 model: the amplitude $\asyn$ (normalized at multipole $80$ and at frequency 23\,GHz), the power spectrum index $\alpha$, the synchrotron spectral index $\betas$, and the dust-synchrotron correlation $\rho$. We show the case of $B$-modes and the Galactic cut $|b|> 10^\circ$. The values correspond to the fifth (i.e. last) multipole bin, $\ell_b=120$.
}\label{fig:pairs}
\end{figure}

Fig.~\ref{fig:pairs} illustrates the correlation structure in the posterior distribution between the main synchrotron parameters, for the particular case of bin 5 (i.e. $\ell_b=120$) and the $B$-modes. 
We find that the amplitude of the power spectrum $\asyn$ is correlated at levels of 0.86 and 0.40 with the 
power spectrum index $\alpha$ and the synchrotron spectral index $\betas$, respectively. However, the correlation with the 
dust-synchrotron correlation parameter ($\rho$) is very small; indeed, the value depicted in the figure of $-0.1$ is the largest one for all bins, with the anti-correlation decreasing down to 0 with decreasing multipoles. 
The values of the $\asyn$–$\alpha$ correlation (the strongest among the model parameters) for all bins, masks and for the two modes are reported in Table~\ref{tab:sfd1}.
This strong correlation between $A_s$ and $\alpha$ is somewhat expected and largely statistical in nature: during the sampling, realizations with a large $A_s$ favour a larger $\alpha$ (flatter SED) to compensate, so that the resulting spectrum remains consistent with the data.

\subsubsection{Dust-synchrotron correlation ($\rho$)}

\begin{figure}
\centerline{\includegraphics[scale=\scfig]{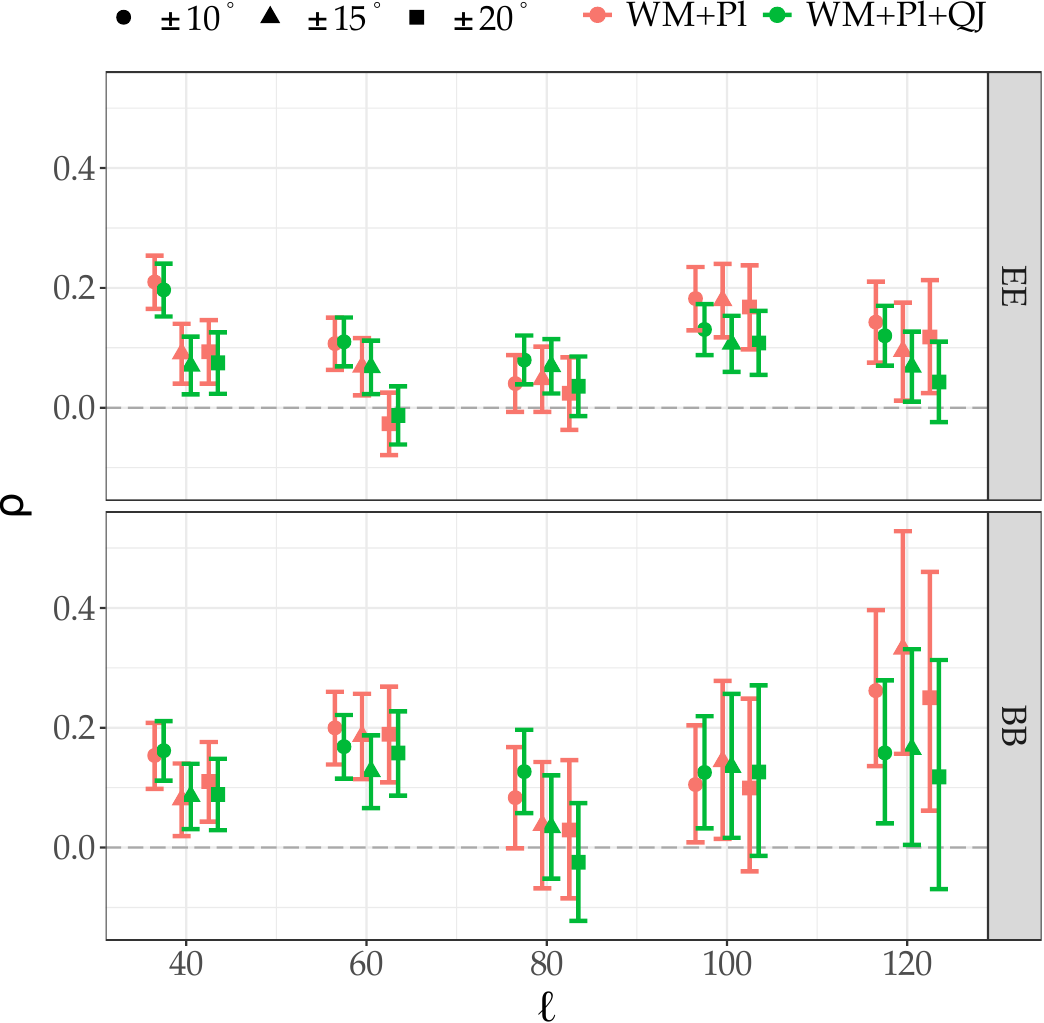}}
\caption{Dust-synchrotron correlation for E-modes (top panel) and B-modes (bottom panel), for the three Galactic cuts and for the two datasets \textit{WM+Pl} (red) and \textit{WM+Pl+QJ} (green). The \textit{NULL} dataset is not shown, as the distribution of the correlation values is nearly uniform between $-1$ and $1$ across all multipole bins. Although all points correspond to the central multipoles $\ell_b = 40, 60, 80, 100$ and $120$, they are slightly shifted horizontally for clarity. Numerical values of $\rho$ can be found in Table~\ref{tab:sfd1}. }\label{fig:rho1}
\end{figure}

Figure~\ref{fig:rho1} shows the inferred dust–synchrotron correlation for the s$\alpha$d1 model, for the three datasets and the three Galactic cuts considered. We find that the correlation remains approximately constant over the multipole range and sky fractions analyzed, with an average value of about $0.15$. 
Our correlation estimates are consistent with the results reported by \citet{Krachmalnicoff:2018imw}, derived from S-PASS and \planck 353\,GHz polarization maps (see their Fig.~12). The small differences can be explained by the variability of the dust-synchrotron correlation across the sky.
Interestingly, the correlation appears to be slightly lower in the $\ell_b = 80$ bin for the $BB$ synchrotron spectrum, where the primordial B-mode signal is expected to peak, consistent with the findings of \citet{Krachmalnicoff:2018imw}. However, more constraining observations are needed to confirm this result.

\subsubsection{Extended synchrotron models: running of the spectral index }
In our analyses, we also tested adding the rolling of the synchrotron SED ($s_{\rm run}$) to the modeling of the synchrotron at the power spectrum level. Since the resulting marginal posterior on $s_{\rm run}$ did not exclude 0, the case with $s_{\rm run}=0$ was used as the default analysis throughout this paper.  However, we discuss this case in Appendix~\ref{app:srun}.

\subsection{Forecasted level of synchrotron contamination for B-mode studies}

\begin{figure}
\centerline{\includegraphics[scale=\scfig]{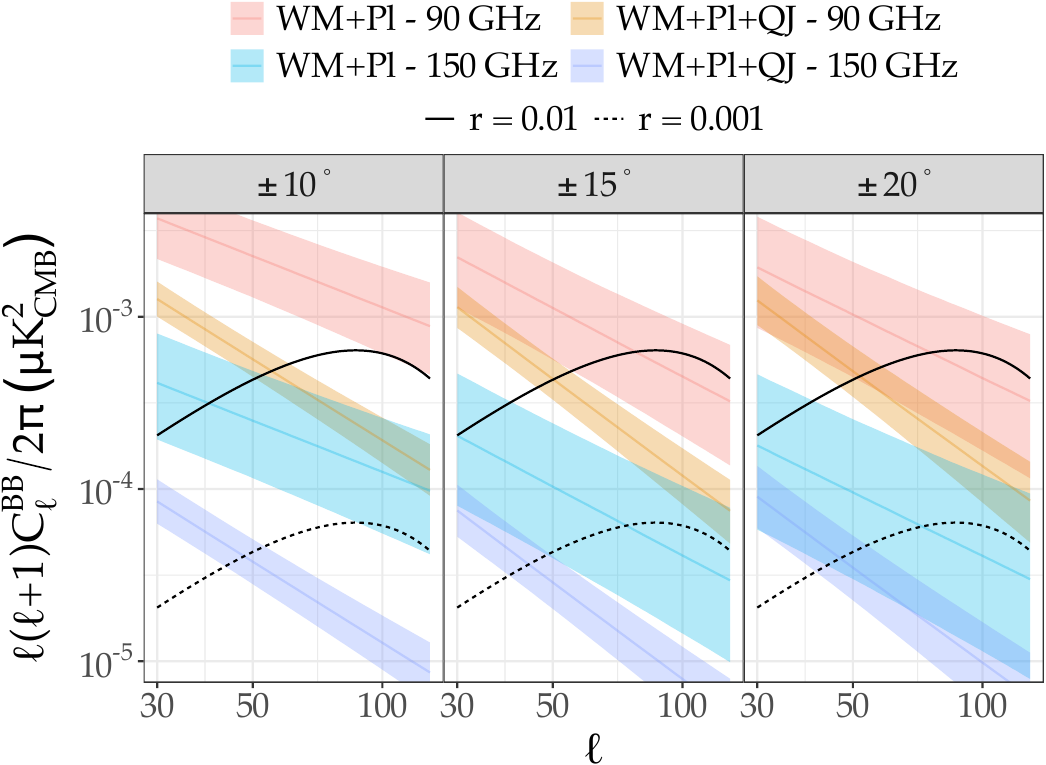}}
\vspace{3mm}
\centerline{\includegraphics[scale=\scfig]{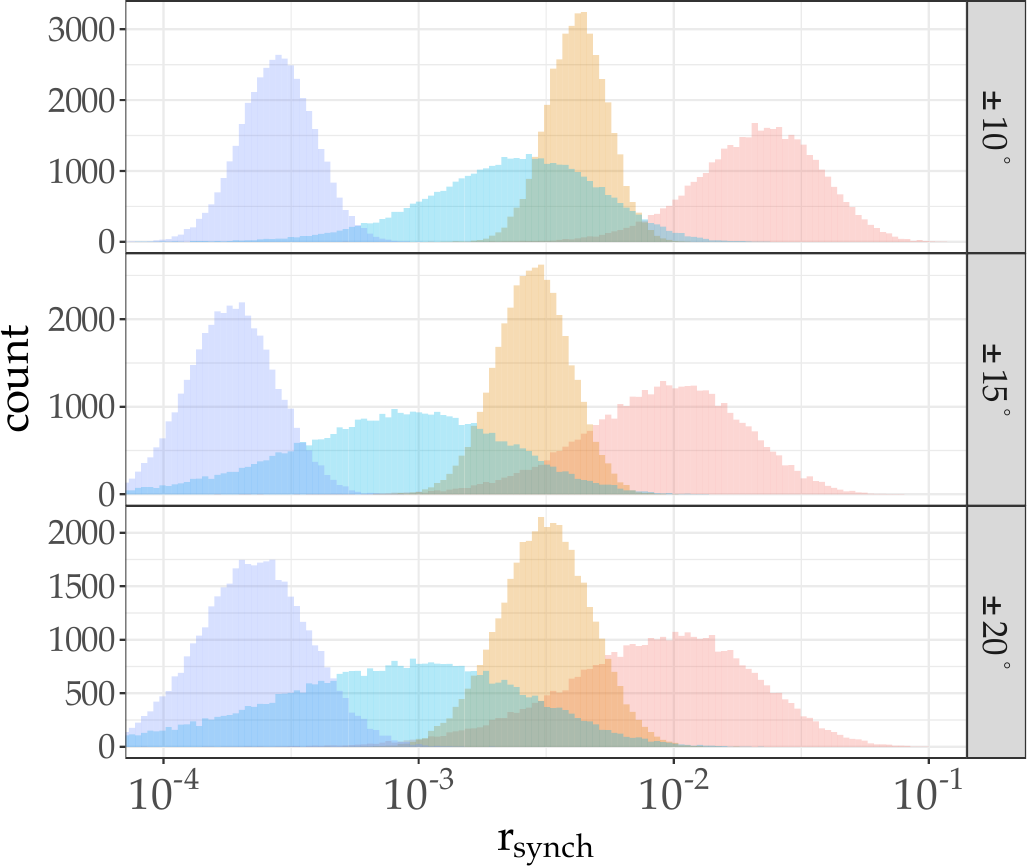}}
\caption{Contamination of the CMB at 90 and 150\,GHz by the synchrotron $B$-modes. Top panel: Power spectra of the synchrotron and the CMB B-modes for two values of the tensor-to-scalar ratio (r=0.01 and r=0.001). The shaded area represent a $1\sigma$ contour, i.e. between the 15.9 and 84.1 quantiles. Bottom panel: Histograms of samples of $r_{\rm synch}$ constructed by injecting the samples of $A_s$ and $\betas$ in Eq.~\ref{eq:rsync}.}
\label{fig:cmbconta}
\end{figure}

\begin{figure}
\centerline{\includegraphics[scale=\scfig]{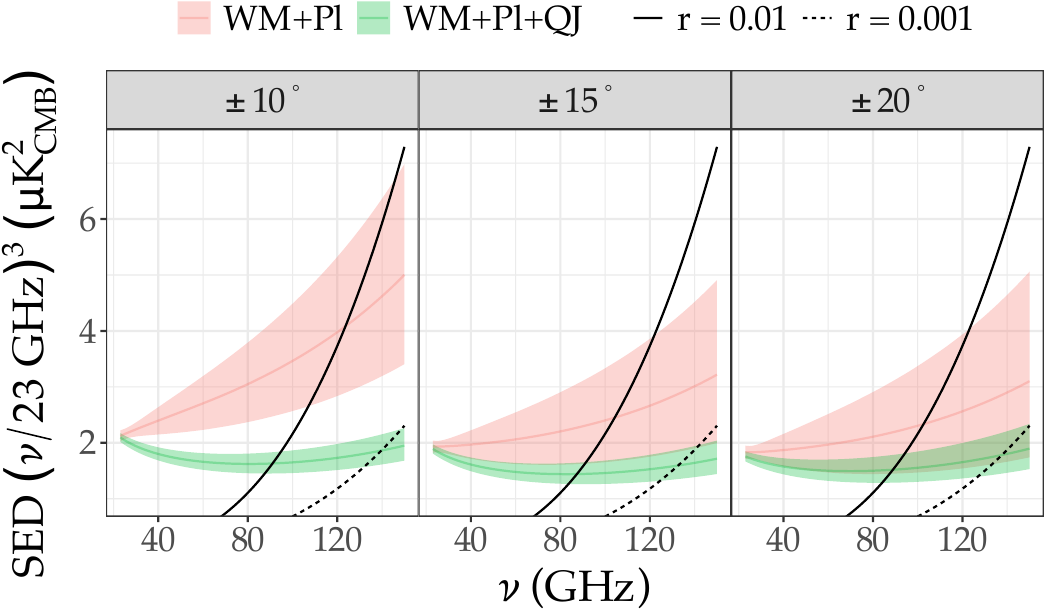}}
\caption{Spectral energy distribution of the standard deviation of the synchrotron $B$-modes, for two sets \textit{WM+Pl} and \textit{WM+Pl+QJ}, and the three galactic cuts. The contours represent the $1\sigma$ contour, i.e. between the 15.9 and 84.1 quantiles of the posterior distribution. The black lines represent the CMB SED for tensor-to-scalar ratio $0.01$ and $0.001$. We note that the SEDs are multiplied by a factor $\nu^3$ for readability purposes.}\label{fig:sed}
\end{figure}

Using the s$\alpha$d1 model and our best-fit results, we can build the distributions of the SED of the synchrotron and of the expected $BB$ power at CMB frequencies most relevant to $B$-mode experiments. The top panel of Fig.~\ref{fig:cmbconta} shows the synchrotron $BB$ power at 90 and 150\,GHz together with the expected CMB $B$-mode amplitude in the cases where the tensor-to-scalar ratio is $r=0.01$ or $r=0.001$. To quantify the level of contamination, we express the power of the synchrotron as an equivalent tensor-to-scalar ratio at the multipole scale of $\ell=80$. Following~\cite{Krachmalnicoff:2018imw}, we define 
\begin{equation}\label{eq:rsync}
    r_{\rm synch}(\nu)=A_s(\nu/23\,\text{GHz})^{2\beta_s}f_\nu^2/A^{\rm CMB}_{\ell=80} \,,
\end{equation}
with $\ell(\ell+1)A^{\rm CMB}_{\ell=80}/2\pi=6.47\times10^{-2}\,\mu K_{\rm CMB}^2$ the amplitude of the unlensed CMB $B$-modes at $\ell=80$ evaluated using the \planck 2018 $\Lambda$CDM best fit model and $r=1$. The factor $f_\nu$ converts amplitudes $\asyn$ from $\mu K_{\rm RJ}$ (antenna temperature) to $\mu K_{\rm CMB}$ (differential CMB thermodynamic temperature), and $f_v=1/a(\nu)$ (with $a(\nu)$ given in Eq.~\ref{eq:KCMB2KRJ}). The bottom panel of Fig.~\ref{fig:cmbconta} shows histograms of $r_{\rm synch}(90\,\text{GHz})$ and $r_{\rm synch}(150\,\text{GHz})$ inferred from both datasets \textit{WM+Pl} and \textit{WM+Pl+QJ}, on the three different masks. The addition of the MFI frequencies reduces the error bars by a factor 2, leading to, for the three masks,
\begin{align}
    r_{\rm synch}(90\,{\rm GHz}) &= \left(4.2_{-1.0}^{+1.3},\, 2.8_{-0.8}^{+1.1},\, 3.1_{-1.1}^{+1.6}\right) \times 10^{-3} \, \\
    r_{\rm synch}(150\,{\rm GHz}) &= \left(2.8_{-0.8}^{+1.1},\, 1.8_{-0.6}^{+0.9},\, 2.3_{-0.9}^{+1.5}\right) \times 10^{-4} \,.
\end{align}
For completeness, Fig.~\ref{fig:sed} shows the SED of the standard deviation at the map level of the synchrotron and CMB $B$-modes. These standard deviations of the synchrotron and the CMB are computed using the same multipole range of the analysis ($30 \le \ell < 130$). As for the spectra we use the fitted spectra for the synchrotron and the \planck best fit spectra so that
\begin{align}
{\rm SED}_{\rm synch.} &= f_\nu\left(\frac{\nu}{\rm 23 GHz}\right)^{\betas}\sqrt{\sum_{\ell=30}^{129} \frac{2\ell+1}{4\pi} \asyn \left(\frac{\ell}{80}\right)^{\alpha}} \label{eq:SED}\\
{\rm SED}_{\rm CMB} &= \sqrt{\sum_{\ell=30}^{129} \frac{2\ell+1}{4\pi} C_\ell^{Planck}} \, .
\end{align}
Fig.~\ref{fig:sed} shows the two cases before and after adding the QUIJOTE frequencies.

Given the current upper limits on $r$, any CMB $B$-mode experiment including frequencies 90~GHz and below targeting intermediate angular scales and tensor-to-scalar ratios below $r = 10^{-2}$ cannot neglect the contribution of synchrotron emission in the sky masks considered in this analysis. This conclusion is consistent with the forecasts presented by \citet{Krachmalnicoff:2018imw} for the southern sky.
QUIJOTE MFI data significantly improve the uncertainty in the estimated synchrotron contamination. It is also worth noting that the slightly steeper derived angular dependence $\alpha$ translates into a lower contamination than in the \textit{WM+Pl} case.

\section{Discussion and conclusions}
\label{sec:ccl}

In this paper, we investigate the properties of polarized synchrotron emission at the power-spectrum level over the multipole range $30 \leq \ell < 130$, using the QUIJOTE MFI wide survey data (10--20\,GHz) in combination with \planck and \wmap. As a main result, we find that the $EE$ and $BB$ power spectra are well described by a power-law model (s$\alpha$d1), with no strong evidence for additional parameters required to capture deviations from this simple behavior. However, the data suggests that the synchrotron SED is flattening from low to high frequencies, but without clear detection, as shown in Appendix~\ref{app:srun}. 

The $EE$ and $BB$ synchrotron power spectra are well fitted by power laws with slopes $\alphaEE = -3.24 \pm 0.12$ and $\alphaBB = -3.81 \pm 0.34$ for the $|b|>20^\circ$ case. 
These values are slightly steeper than those reported by \citet{Martire2022}, but are consistent with those of \citet{Krachmalnicoff:2018imw}. The inclusion of QUIJOTE MFI data suggests a steepening of the $BB$ spectrum relative to $EE$, although the significance of this finding remains at the $\sim 2\sigma$ level. Future analyses with improved sensitivity at MFI frequencies \citep{MFI2,mfi2_synch_forecast} or adding lower frequency data \citep[e.g. C-BASS][]{Jones:2018lvt} will help to clarify this trend.

The best-fit synchrotron spectral index is found to be $\betas^{\rm EE} = -3.26\pm0.05$ and $\betas^{\rm BB} = -3.25\pm0.10$ for the $|b|>20^\circ$. These values are consistent with previous analyses based on real-space statistics using QUIJOTE-MFI data in combination with \planck and \wmap \citep{MFIcompsep_pol,semiblind}. 
We find no evidence for spatial variations of the spectral index $\betas$ in any of the three Galactic cuts considered in this work ($|b|>10^\circ$, $15^\circ$, and $20^\circ$). A more detailed analysis, based on smaller sky patches to estimate the synchrotron power spectrum, is deferred to future work. We also note that the level of spatial variations reported in real-space analyses (0.13 from {\citealt{MFIcompsep_pol}}) 
is comparable to the uncertainties on $\betas$ for the $BB$ spectra; therefore, any detectable variations with the current datasets would be expected primarily in the $EE$ spectra.
When compared with results based on \planck and \wmap data alone, we find that the inclusion of QUIJOTE–MFI data yields a marginally steeper value of $\betas$, although with low statistical significance (at the $\sim 2.4\sigma$ level for the $|b|>10^\circ$ case).

We find a synchrotron $BB$/$EE$ ratio of approximately $0.2$, consistent with previous analyses based on \planck and \wmap data \citep{Martire2022} and on QUIJOTE–MFI observations \citep{mfiwidesurvey}. These values are significantly lower than those reported by \citet{Krachmalnicoff:2018imw} using S-PASS data at 2.3\,GHz, likely due to residual Faraday rotation in those maps.
Overall, these results indicate again that the synchrotron $BB$/$EE$ ratios are lower than those measured for thermal dust, which are typically of the order of 0.5 \citep{Adam:2014bub, Planck2018-xi} and around 0.7 specifically on the masks we used for the analyses of this paper.
Regarding the spatial correlation with dust, we find that the cross-correlation for both $E$- and $B$-modes lies in the range 0.1--0.25 over $30 \leq \ell \leq 300$, consistent with \citet{Krachmalnicoff:2018imw}.

Using the previous results for the synchrotron spectra, we have investigated the level of synchrotron contamination in future CMB $B$-mode measurements. For the sky region and multipole range considered in this analysis, we find that synchrotron emission is, on average, equivalent to $r\sim3\times10^{-3}$ at 90\,GHz and to $r\sim 2 \times 10^{-4}$ at 150\,GHz, at the angular scales relevant for primordial signal detection. 
Current and future experiments aim at least at a $2\sigma$ detection of $r=0.01$. Therefore, any analysis targeting primordial $B$-mode signals must include a mitigation strategy to account for synchrotron emission.

In this study, we neglect Faraday depolarization in the data modelling. This is a valid assumption for \wmap and \planck, and also for QUIJOTE–MFI in the sky regions considered here, as the standard deviation of the difference between QUIJOTE and \planck polarization angles over the adopted masks is about $0.6^\circ$, with no evidence of Faraday rotation \citep[see][]{mfiwidesurvey}.

We assume that only synchrotron and dust emission contribute to the observed polarization. The fact that the single-component model (s$\alpha$d1) is preferred by the data for the synchrotron emission indicates that no additional polarized component, in particular AME, is detected at the power-spectrum level.
A dedicated study is required to place quantitative upper limits on the AME polarization fraction. This can be achieved by extending the data model to include an additional component with an unconstrained SED. Such limits could then be derived from the posterior distribution of the power of this extra component at each frequency. This is left for future work.

\section*{Data availability}
The QUIJOTE maps used in this work are publicly available as part of the QUIJOTE-MFI Data Release 1 (DR1) at \url{https://doi.org/10.26698/quijote-mfi-dr1}, and can also be downloaded from the QUIJOTE website\footnote{QUIJOTE web: \url{https://research.iac.es/proyecto/quijote/}}.

\section*{Acknowledgements}

We thank the staff of the Teide Observatory for invaluable assistance
in the commissioning and operation of QUIJOTE.
The {\it QUIJOTE} experiment is being developed by the Instituto de
Astrofisica de Canarias (IAC), the Instituto de Fisica de Cantabria
(IFCA), and the Universities of Cantabria, Manchester and Cambridge.
Partial financial support was provided by the Spanish Ministry of Science and
Innovation (MICIU/AEI/10.13039/501100011033) through the projects
AYA2007-68058-C03- 01, AYA2007-68058-C03-02, 
AYA2010-21766-C03-01, AYA2010-21766-C03- 02, AYA2014-60438-P, 
ESP2015-70646-C2-1-R, AYA2017-84185-P, ESP2017- 83921-C2-1-R, 
PGC2018-101814-B-I00, PID2019-110610RB-C21, PID2020-120514GB-I00, 
IACA13-3E-2336, IACA15-BE-3707, EQC2018-004918-P, PID2022-139223OB-C21,
PID2023-150398NB-I00 and PID2023-151567NB-I00; 
the Red de Investigaci\'on RED2022-134715-T; 
the Severo Ochoa Programs SEV-2015-0548, CEX2019-000920-S and
CEX2025-001609-S;  the Maria de Maeztu Programs MDM-2017-0765 and
CEX2025-001620-M; and the Consolider-Ingenio project CSD2010-00064
(EPI: Exploring the Physics of Inflation).
We acknowledge support from the ACIISI, Consejeria de Economia,
Conocimiento y Empleo del Gobierno de Canarias and the European
Regional Development Fund (ERDF) under grant ProID2020010108. 
This project has received funding from the European Union's Horizon
2020 research and innovation programme under grant agreement
No. 687312 (RADIOFOREGROUNDS), and from the Horizon Europe research
and innovation programme under grant agreement No. 101135036
(RadioForegroundsPlus).

\bibliographystyle{mnras}
\bibliography{biblio,quijote}



\appendix

\section{Derived synchrotron properties for the s1d1 model}
\label{app:s1d1}

Although the s$\alpha$d1 model is favored by the Bayes factors (as discussed in the main text, see Sect.~\ref{sec:modelsel}), we use this appendix to illustrate that the main synchrotron parameters, such as the spectral index, are robust with respect to the adopted model. To this end, we present here the parameters derived using the second-most favored model according to the Bayes factor, namely s1d1. Table~\ref{tab:synchrotron_s1d1} lists the synchrotron-related parameters, namely the spectral index and the amplitude, which can be compared to the values reported in Table~\ref{tab:sfd1} in the main text. Note that in this model, the synchrotron amplitude is fitted independently in each multipole bin.
For completeness, Table~\ref{tab:dust_s1d1} presents the corresponding dust-related parameters, which can be compared with those reported in appendix~\ref{app:dust_sfd1}, showing also a good consistency with those derived from the s$\alpha$d1 model.

\begin{table*}
    \centering
    
    \caption{Values of the synchrotron related parameters when inferring using the s1d1 model. Synchrotron amplitude in each bin is computed at the pivot frequency of 22.8\,GHz, and the units are $\times 10^{-3}$\,$\mu$K$_{\rm RJ}^2$. The values are given as $x\pm y$ with $x$ the median of the marginal posterior distribution and $y$ the interquantile range between quantiles 0.159 and 0.841 divided by two.}
    \begin{tabular}{ccccccccc}
    \hline 
Mask & Dataset & mode & $\beta_s$ & $A_{s,40}$ & $A_{s,60}$ & $A_{s,80}$ & $A_{s,100}$ & $A_{s,120}$ \\ 
 \hline
$\pm10^\circ$ & WM+Pl & EE & $-3.00\pm0.09$ & $94.2\pm6.9$ & $24.9\pm1.9$ & $10.0\pm1.0$ & $5.80\pm0.79$ & $2.06\pm0.70$ \\ 
 &  & BB & $-2.9\pm0.2$ & $18.9\pm2.1$ & $6.05\pm0.90$ & $0.69\pm0.67$ & $1.84\pm0.64$ & $0.44\pm0.48$ \\ 
 & WM+Pl+QJ & EE & $-3.22\pm0.03$ & $99.3\pm6.8$ & $25.4\pm1.7$ & $10.47\pm0.80$ & $5.67\pm0.51$ & $2.78\pm0.39$ \\ 
 &  & BB & $-3.33\pm0.07$ & $21.6\pm1.9$ & $4.95\pm0.60$ & $1.61\pm0.37$ & $0.47\pm0.37$ & $0.29\pm0.31$ \\ 
$\pm15^\circ$ & WM+Pl & EE & $-3.1\pm0.1$ & $82.5\pm6.5$ & $20.8\pm1.8$ & $8.1\pm1.0$ & $5.16\pm0.82$ & $0.95\pm0.80$ \\ 
 &  & BB & $-3.1\pm0.3$ & $17.1\pm2.1$ & $5.06\pm0.92$ & $0.13\pm0.34$ & $1.44\pm0.76$ & $0.17\pm0.27$ \\ 
 & WM+Pl+QJ & EE & $-3.24\pm0.03$ & $86.7\pm6.4$ & $21.3\pm1.6$ & $8.62\pm0.77$ & $5.08\pm0.52$ & $2.15\pm0.39$ \\ 
 &  & BB & $-3.34\pm0.08$ & $18.7\pm1.8$ & $3.98\pm0.57$ & $1.02\pm0.38$ & $0.11\pm0.22$ & $0.082\pm0.180$ \\ 
$\pm20^\circ$ & WM+Pl & EE & $-3.1\pm0.1$ & $78.8\pm6.7$ & $18.9\pm1.8$ & $7.5\pm1.1$ & $4.55\pm0.87$ & $0.50\pm0.66$ \\ 
 &  & BB & $-3.1\pm0.3$ & $15.1\pm2.1$ & $4.32\pm0.94$ & $0.23\pm0.48$ & $1.34\pm0.85$ & $0.15\pm0.29$ \\ 
 & WM+Pl+QJ & EE & $-3.26\pm0.03$ & $82.3\pm6.5$ & $19.0\pm1.5$ & $7.85\pm0.77$ & $4.39\pm0.52$ & $1.63\pm0.40$ \\ 
 &  & BB & $-3.25\pm0.10$ & $16.3\pm1.8$ & $3.24\pm0.58$ & $0.95\pm0.44$ & $0.13\pm0.25$ & $0.11\pm0.24$ \\
\hline
    \end{tabular}
    
    \label{tab:synchrotron_s1d1}
    \end{table*}
    
\begin{table*}
    \centering
    
    \caption{Values of the dust related parameters when inferring using the s1d1 model. Dust amplitude in each bin is computed at the pivot frequency of 353\,GHz, and the units are $\times 10^{-3}$\,$\mu$K$_{\rm RJ}^2$. The values are given as $x\pm y$ with $x$ the median of the marginal posterior distribution and $y$ the interquantile range between quantiles 0.159 and 0.841 divided by two.}
    \begin{tabular}{ccccccccc}
    \hline 
Mask & Dataset & mode & $\beta_d$ & $A_{d,40}$ & $A_{d,60}$ & $A_{d,80}$ & $A_{d,100}$ & $A_{d,120}$ \\ 
 \hline
$\pm10^\circ$ & WM+Pl & EE & $1.497\pm0.007$ & $20.5\pm1.3$ & $7.07\pm0.37$ & $4.31\pm0.19$ & $2.445\pm0.099$ & $1.515\pm0.057$ \\ 
 &  & BB & $1.504\pm0.006$ & $14.42\pm0.92$ & $5.75\pm0.30$ & $3.03\pm0.14$ & $1.594\pm0.065$ & $1.251\pm0.047$ \\ 
 & WM+Pl+QJ & EE & $1.497\pm0.007$ & $20.5\pm1.3$ & $7.07\pm0.36$ & $4.31\pm0.19$ & $2.44\pm0.10$ & $1.514\pm0.058$ \\ 
 &  & BB & $1.502\pm0.006$ & $14.43\pm0.92$ & $5.75\pm0.30$ & $3.03\pm0.14$ & $1.594\pm0.065$ & $1.252\pm0.047$ \\ 
$\pm15^\circ$ & WM+Pl & EE & $1.51\pm0.01$ & $10.03\pm0.67$ & $4.40\pm0.24$ & $2.36\pm0.11$ & $1.557\pm0.068$ & $0.903\pm0.036$ \\ 
 &  & BB & $1.515\pm0.008$ & $7.96\pm0.53$ & $3.58\pm0.20$ & $2.056\pm0.099$ & $1.006\pm0.044$ & $0.692\pm0.028$ \\ 
 & WM+Pl+QJ & EE & $1.51\pm0.01$ & $10.02\pm0.68$ & $4.40\pm0.24$ & $2.36\pm0.11$ & $1.556\pm0.069$ & $0.903\pm0.037$ \\ 
 &  & BB & $1.514\pm0.008$ & $7.96\pm0.54$ & $3.58\pm0.20$ & $2.056\pm0.098$ & $1.006\pm0.044$ & $0.692\pm0.028$ \\ 
$\pm20^\circ$ & WM+Pl & EE & $1.49\pm0.01$ & $7.65\pm0.55$ & $2.81\pm0.17$ & $1.496\pm0.078$ & $1.088\pm0.051$ & $0.606\pm0.027$ \\ 
 &  & BB & $1.52\pm0.01$ & $5.41\pm0.39$ & $2.33\pm0.14$ & $1.294\pm0.067$ & $0.631\pm0.030$ & $0.399\pm0.018$ \\ 
 & WM+Pl+QJ & EE & $1.49\pm0.01$ & $7.64\pm0.55$ & $2.81\pm0.17$ & $1.496\pm0.077$ & $1.088\pm0.051$ & $0.606\pm0.027$ \\ 
 &  & BB & $1.52\pm0.01$ & $5.41\pm0.39$ & $2.33\pm0.14$ & $1.293\pm0.067$ & $0.630\pm0.030$ & $0.400\pm0.018$ \\
\hline
    \end{tabular}
    
    \label{tab:dust_s1d1}
    \end{table*}

\section{Derived dust properties for the s$\alpha$d1 model}
\label{app:dust_sfd1}

Table~\ref{tab:dust_sfd1} presents the derived dust parameters (spectral index $\betad$, and the amplitude $\adust$ of the dust power spectra normalized at 353\,GHz for each of the different multipole bins) for the same s$\alpha$d1 model discussed in 
Sect~\ref{sec:results}. 
Although they are not the primary goal of this paper, we note that the retrieved $\betad$ parameters are fully consistent with the average values derived in previous works \citep[see e.g. tables C.2 and C.3 in][]{Planck2018-xi}. 
The amplitude and $BB/EE$ ratio values cannot be directly compared, as they are evaluated over different sky masks. However, we note that the average $BB/EE$ ratio for thermal dust within the three masks lies in the range $0.7$--$0.8$, significantly higher than the values reported for synchrotron, as reported in the main text. 

\begin{table*}
    \centering
    
    \caption{Values of the dust related parameters when inferring using the s$\alpha$d1 model. Dust amplitude in each bin is computed at the pivot frequency of 353\,GHz, and the units are $\times 10^{-3}$\,$\mu$K$_{\rm RJ}^2$. The values are given as $x\pm y$ with $x$ the median of the marginal posterior distribution and $y$ the interquantile range between quantiles 0.159 and 0.841 divided by two.}
    \begin{tabular}{ccccccccc}
    \hline 
Mask & Dataset & mode & $\beta_d$ & $A_{d,40}$ & $A_{d,60}$ & $A_{d,80}$ & $A_{d,100}$ & $A_{d,120}$ \\ 
 \hline
$\pm10^\circ$ & WM+Pl & EE & $1.497\pm0.007$ & $20.5\pm1.3$ & $7.07\pm0.37$ & $4.31\pm0.20$ & $2.443\pm0.099$ & $1.515\pm0.057$ \\ 
 &  & BB & $1.504\pm0.006$ & $14.41\pm0.91$ & $5.75\pm0.30$ & $3.03\pm0.14$ & $1.594\pm0.065$ & $1.252\pm0.047$ \\ 
 & WM+Pl+QJ & EE & $1.497\pm0.007$ & $20.5\pm1.3$ & $7.08\pm0.37$ & $4.31\pm0.19$ & $2.441\pm0.099$ & $1.514\pm0.057$ \\ 
 &  & BB & $1.502\pm0.006$ & $14.42\pm0.92$ & $5.75\pm0.30$ & $3.03\pm0.14$ & $1.594\pm0.065$ & $1.251\pm0.047$ \\ 
$\pm15^\circ$ & WM+Pl & EE & $1.51\pm0.01$ & $10.02\pm0.67$ & $4.40\pm0.24$ & $2.35\pm0.11$ & $1.557\pm0.068$ & $0.903\pm0.037$ \\ 
 &  & BB & $1.515\pm0.008$ & $7.97\pm0.54$ & $3.58\pm0.20$ & $2.056\pm0.099$ & $1.006\pm0.044$ & $0.693\pm0.028$ \\ 
 & WM+Pl+QJ & EE & $1.51\pm0.01$ & $10.02\pm0.67$ & $4.40\pm0.24$ & $2.36\pm0.11$ & $1.556\pm0.067$ & $0.903\pm0.037$ \\ 
 &  & BB & $1.514\pm0.008$ & $7.97\pm0.54$ & $3.58\pm0.20$ & $2.055\pm0.099$ & $1.006\pm0.044$ & $0.693\pm0.028$ \\ 
$\pm20^\circ$ & WM+Pl & EE & $1.49\pm0.01$ & $7.64\pm0.55$ & $2.81\pm0.17$ & $1.496\pm0.078$ & $1.088\pm0.051$ & $0.607\pm0.027$ \\ 
 &  & BB & $1.52\pm0.01$ & $5.42\pm0.39$ & $2.33\pm0.14$ & $1.294\pm0.067$ & $0.630\pm0.030$ & $0.400\pm0.018$ \\ 
 & WM+Pl+QJ & EE & $1.49\pm0.01$ & $7.64\pm0.55$ & $2.81\pm0.17$ & $1.496\pm0.078$ & $1.087\pm0.051$ & $0.606\pm0.027$ \\ 
 &  & BB & $1.52\pm0.01$ & $5.41\pm0.38$ & $2.33\pm0.14$ & $1.294\pm0.068$ & $0.630\pm0.030$ & $0.400\pm0.018$ \\
\hline
    \end{tabular}
    
    \label{tab:dust_sfd1}
    \end{table*}

\section{Constraints on the rolling of the synchrotron SED}
\label{app:srun}

We tested adding the rolling of the synchrotron SED ($\srun$) to the modeling of the synchrotron. The $\srun$ parameter models a potential curvature of the synchrotron SED as 
\begin{equation}
    A_s(\nu) \propto (\nu/\nu_s)^{\betas+\srun\log(\nu/\nu_s)} \,,
\end{equation}
with $\nu_s=23$ GHz. To constrain $\srun$, we introduced an alternative model ssd1 and fitted it to the \textit{WM+Pl+QJ} dataset on the three masks. The ssd1 model has $\srun$ in its set of parameters $\theta$ in addition to all the parameters of the s$\alpha$d1 model. The prior on the $\srun$ parameter is a Gaussian distribution with mean 0 and standard deviation 0.1. The rest of the parameters have the same prior distributions. The choice of the prior distribution on $\srun$ is motivated by the existing constraints found in the literature, e.g.
\cite{Hensley:2021ydb} and \cite{Krachmalnicoff:2018imw} that analyze synchrotron properties between 27 and 280\,GHz with reference frequency $\nu_s=23$ GHz and between 2.3 and 23\,GHz with reference frequency 9\,GHz, respectively, and that found mean values of $\srun$ compatible with zero with standard deviation of 0.01 and 0.1, respectively.
Table~\ref{tab:srun} presents the results of the fit. For all combinations of mask and polarization mode, the fitted value for $\srun$ is compatible with 0 with characteristic uncertainties at the level of $0.05$--$0.1$ for the different sky masks, and the rest of the parameters have very similar posterior distribution as those in the s$\alpha$d1 model. The variance of the marginal posterior distribution of $\srun$ The strongest constraint on the $\srun$ parameter is for the $E$-modes with $|\srun| < 0.21$ at the $3\sigma$ level. The model selection based on Bayes factors puts the two models s$\alpha$d1 and ssd1 at the same level, i.e. none of them stands out as favored by the data. Based on these results, we adopted $\srun = 0$ as the baseline model in the main text. 

\begin{table}
\centering
\caption{Values of $s_{\rm run}$ for the ssd1 model on the three masks using the \textit{WM+Pl+QJ} dataset.}
\begin{tabular}{cccc}
\hline 
Mode & $\pm10^\circ$ & $\pm15^\circ$ & $\pm20^\circ$ \\ 
 \hline
$EE$ & $0.07\pm0.05$ & $0.03\pm0.06$ & $0.003\pm0.070$ \\ 
$BB$ & $0.10\pm0.09$ & $0.05\pm0.09$ & $0.04\pm0.10$ \\
\hline
\end{tabular}

\label{tab:srun}
\end{table}

\end{document}